\documentclass[aps,pre,reprint,floatfix,longbibliography,superscriptaddress,amsmath,amssymb,compress]{revtex4-1}
\pdfoutput=1

\setcitestyle{numbers,square,comma}

\usepackage{graphicx}
\usepackage{color}
\usepackage{dsfont}

\usepackage{booktabs}

\usepackage{siunitx}

\AtBeginDocument{
  \heavyrulewidth=.08em
  \lightrulewidth=.06em
  \cmidrulewidth=.03em
  \belowrulesep=.65ex
  \belowbottomsep=0pt
  \aboverulesep=.4ex
  \abovetopsep=0pt
  \cmidrulesep=\doublerulesep
  \cmidrulekern=.5em
  \defaultaddspace=.5em
  }

\newcommand{\papertitle}{Non-dispersive one-way signal amplification in sonic metamaterials}

\newcommand{\subfiglabel}[1]{{\bfseries #1}}

\newcommand{\figref}[1]{Fig.~\ref{#1}}
\newcommand{\figureref}[1]{Figure~\ref{#1}}
\newcommand{\subfigref}[2]{\figref{#1}\subfiglabel{#2}} 
\newcommand{\subfigureref}[2]{\figureref{#1}\subfiglabel{#2}}  

\newcommand{\appref}[1]{Appendix~\ref{#1}}
\newcommand{\eqnref}[1]{Eq.~\eqref{#1}}
\newcommand{\equationref}[1]{Equation~\eqref{#1}}

\newcommand{\Oms}{\Omega_\text{s}}

\newcommand{\acktext}{This work was informed by preliminary analyses conducted
by Nathan Villiger, Maxx Miller, and Pragalv Karki on related systems. We thank
Abhijeet Melkani for useful discussions and feedback on the manuscript. Work was
partially supported by the National Science Foundation under award
CMMI-2128671.}

\begin{document}
\title{\papertitle}
\author{Noah Kruss}
\affiliation{Department of Physics, University of Oregon, Eugene, OR 97403}
\author{Jayson Paulose}
\affiliation{Department of Physics, University of Oregon, Eugene, OR 97403}
\affiliation{Institute for Fundamental Science and Materials Science Institute,  University of Oregon, Eugene, OR
  97403}

\begin{abstract}

Parametric amplification---injecting energy into waves via periodic modulation
of system parameters---is typically restricted to specific multiples of the
modulation frequency. However, broadband parametric amplification can be
achieved in active metamaterials which allow local parameters to be modulated
both in space and in time. Inspired by the concept of luminal metamaterials in
optics, we describe a mechanism for one-way amplification of sound waves across
an entire frequency band using spacetime-periodic modulation of local
stiffnesses in the form of a traveling wave. When the speed of the modulation
wave approaches that of the speed of sound in the metamaterial---a regime called
the sonic limit---nearly all modes in the forward-propagating acoustic band are
amplified, whereas no amplification occurs in the reverse-propagating band. To
eliminate divergences that are inherent to the sonic limit in continuum
materials, we use an exact Floquet-Bloch approach to compute the dynamic
excitation bands of discrete periodic systems. We find wide ranges of parameters
for which the amplification is nearly uniform across the lowest-frequency band,
enabling amplification of wavepackets while preserving their speed, shape, and
spectral content. Our mechanism provides a route to designing acoustic
metamaterials which can propagate wave pulses without losses or distortion
across a wide range of frequencies.

\end{abstract}

\maketitle

\section{Introduction}

Parametric amplification---feeding energy into oscillatory modes through a
periodic modulation of the underlying stiffness or coupling
parameters---provides a technologically-relevant route to boosting signals and
overcoming losses in electromagnetic~\cite{Cullen1958,Tien1958},
optical~\cite{Baumgartner1979} and mechanical~\cite{Rugar1991} systems.
Typically, parametric amplification occurs only for a discrete set of modes
which satisfy specific frequency relationships with the modulation
frequency~\cite{landau1982mechanics}, which obstructs its use to amplify
propagating signals with multiple frequency components such as localized
wavepackets. However, when the parameter modulation is itself a traveling wave
through the medium, interference effects enable amplification over a wide range
of signal frequencies with a single modulation
frequency~\cite{Cassedy1963,Cassedy1967}, opening up possibilities for
amplification and loss-compensation of multispectral signals as long as the
desired spacetime parameter modulation can be achieved.

Active metamaterials---artificial structures whose properties can be modulated
using external fields~\cite{Boardman2011,Zangeneh-Nejad2019,Wang2020}---provide
a promising platform for broadband parametric amplification using traveling
waves~\cite{Galiffi2019}.
In the realm of acoustics, traveling-wave modulation
of elastic stiffnesses has primarily been used to achieve nonreciprocal
transport~\cite{Wang2018,Trainiti2019,Yi2019,Chen2019b,Attarzadeh2020,Marconi2020,Xia2021},
although parametric amplification has also been
observed albeit in narrow frequency ranges~\cite{Trainiti2019}. Despite rapid developments in active acoustic
metamaterial platforms which enable on-demand spatiotemporal modulation of
acoustic parameters across a wide range of length and frequency
scales~\cite{Zangeneh-Nejad2019,Nassar2020}, traveling-wave parametric
amplification remains unexploited as a mechanism to boost multispectral signals
in active acoustic metamaterials.

\begin{figure}
  \centering
  \includegraphics{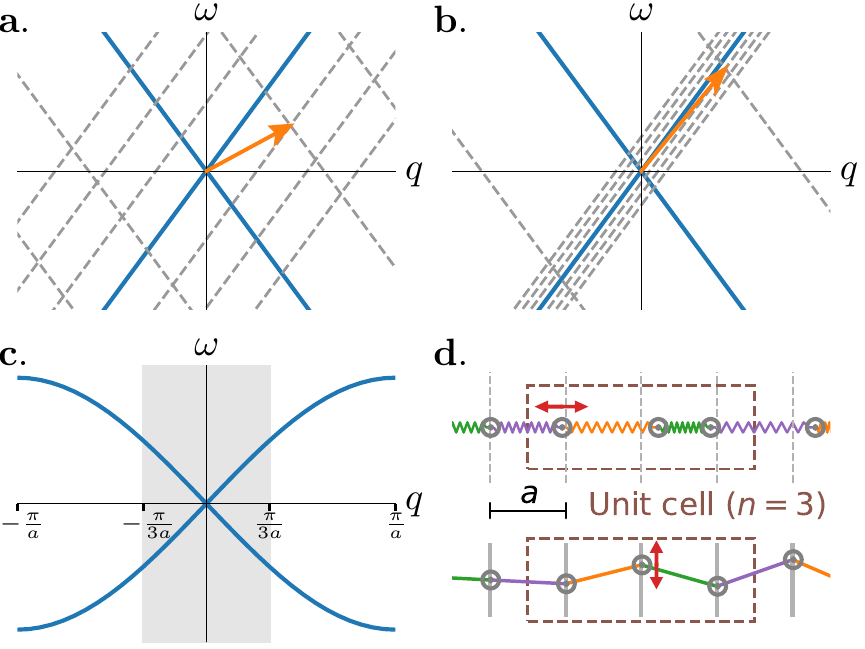}
  \caption{{\bfseries Sonic metamaterials.}
    \subfiglabel{a}, A 1D continuum system displays a linear dispersion relation
    for sound waves (solid lines). When the local stiffness $k(x,t)$ is modulated
    according to a traveling wave, $k(x,t) = k_0[1+\delta\cos(gx-\Omega t)]$, the
    resulting normal modes arise from the coupling of the unmodulated modes with
    their harmonic replicas (dashed lines) displaced by integer multiples of the
    vector $(g,\Omega)$ (arrow) in the frequency-wavevector plane. Here, $\Omega
    /g = 0.4 v$.     
    \subfiglabel{b}, Same as \subfiglabel{a} with $\Omega/g = 0.9v$. The
    first six harmonics are shown. As $\Omega/g \to v$, all harmonics overlap
    along $\omega = vq$.
    \subfiglabel{c}, A periodic spring-mass chain with lattice spacing $a$ has a
    nonlinear dispersion relation across the Brillouin zone. Shaded region
    shows the reduced Brillouin zone when spring constants are modulated
    as shown in \subfiglabel{d}.  
    \subfiglabel{d}, (top) Section of an infinite spring chain with lattice
    spacing $a$; horizontal
    displacements (arrow) from equilibrium (dashed vertical lines) are coupled by a
    repeating set of three unique
    spring constants (colors) as defined by \eqnref{eq:travelingwave} with
    $n=3$. (bottom) The same equation of motion is obeyed by vertical displacements of
    masses restricted to slide along immobile rigid bars spaced by $a$ (solid vertical lines) and
    coupled by tensed strings whose tensions (colors) are modulated according to
    \eqnref{eq:travelingwave}. 
  }
  \label{fig:intro}
\end{figure}

Here, we show that a traveling-wave stiffness modulation can generate broadband
parametric amplification in acoustic systems as a consequence of instabilities
that arise when the speed of the traveling-wave modulation is close to the speed
of sound in the medium~\cite{Roe1959,Simon1960,Hessel1961}---a situation termed
the \emph{sonic limit}~\cite{Cassedy1963}. In the sonic limit, approximate
techniques such as coupled-mode theory and plane-wave expansions, commonly used
to compute the response of time-modulated
metamaterials~\cite{Swinteck2015,Trainiti2016,NassarH.2017,Wang2018a}, are known
to break down~\cite{Hessel1961,Cassedy1963}. Instead, we develop a Floquet-Bloch
technique to calculate the exact dispersion relation of a discrete system of
masses connected by springs with spacetime-modulated stiffnesses. We find that
the acoustic gain (the imaginary part of the complex frequency) can be made
nearly constant over a broad range of frequencies and quasimomenta, allowing
coherent amplification and loss-mitigation of acoustic signals with a broad
spectral content. The gain is controlled by the modulation strength, which allows our
technique to be dynamically tuned to produce the desired amplification, or to
finely balance losses for unattenuated sound transmission. It is also strongly
directional, allowing highly non-reciprocal response with amplified transport of
signals in one direction and strong attenuation in the opposite direction. As a
technologically-relevant illustration of our approach, we demonstrate
dispersion-free amplification and loss-compensation of propagating wave pulses
in modulated spring-mass chains.

The physical mechanism underlying the sonic limit is illustrated in
\figref{fig:intro} for a continuum one-dimensional (1D) system which admits a
linear dispersion relation $\omega(q) = \pm v q$ between the inverse wavelength,
or quasimomentum, $q$ of traveling waves and their oscillation frequency
$\omega$ when the underlying stiffness
constants are uniform~\cite{Cassedy1963,Galiffi2019}. If the stiffness is
perturbed by a periodic traveling-wave modulation with quasimomentum $g$ and
frequency $\Omega$, Floquet-Bloch theory dictates that the original normal modes
become strongly coupled with harmonics that are displaced by integer multiples
of the vector $(g,\Omega)$ on the quasimomentum-frequency plane
(\subfigref{fig:intro}{a}).
When $\Omega / g$ approaches the speed of sound $v$, all harmonics of the
original set of modes begin to overlap along the branch $\omega = vq$
(\subfigref{fig:intro}{b}), signifying a pile-up of harmonic contributions at
the sonic limit of the modulated medium. Because of these
overlapping contributions from a technically infinite set of harmonics,
calculations of the dispersion relation of the infinite continuum system do not
converge~\cite{Cassedy1963}. However, the response of a finite system over finite time intervals
can still be computed, and has been shown to exhibit broadband amplification and
high-frequency harmonic generation in optical metamaterials where the analogous
situation has been termed the \emph{luminal} limit~\cite{Galiffi2019}.

\section{Floquet-Bloch band structures of time-modulated spring networks}

\begin{figure*}[t]
  \centering
  \includegraphics{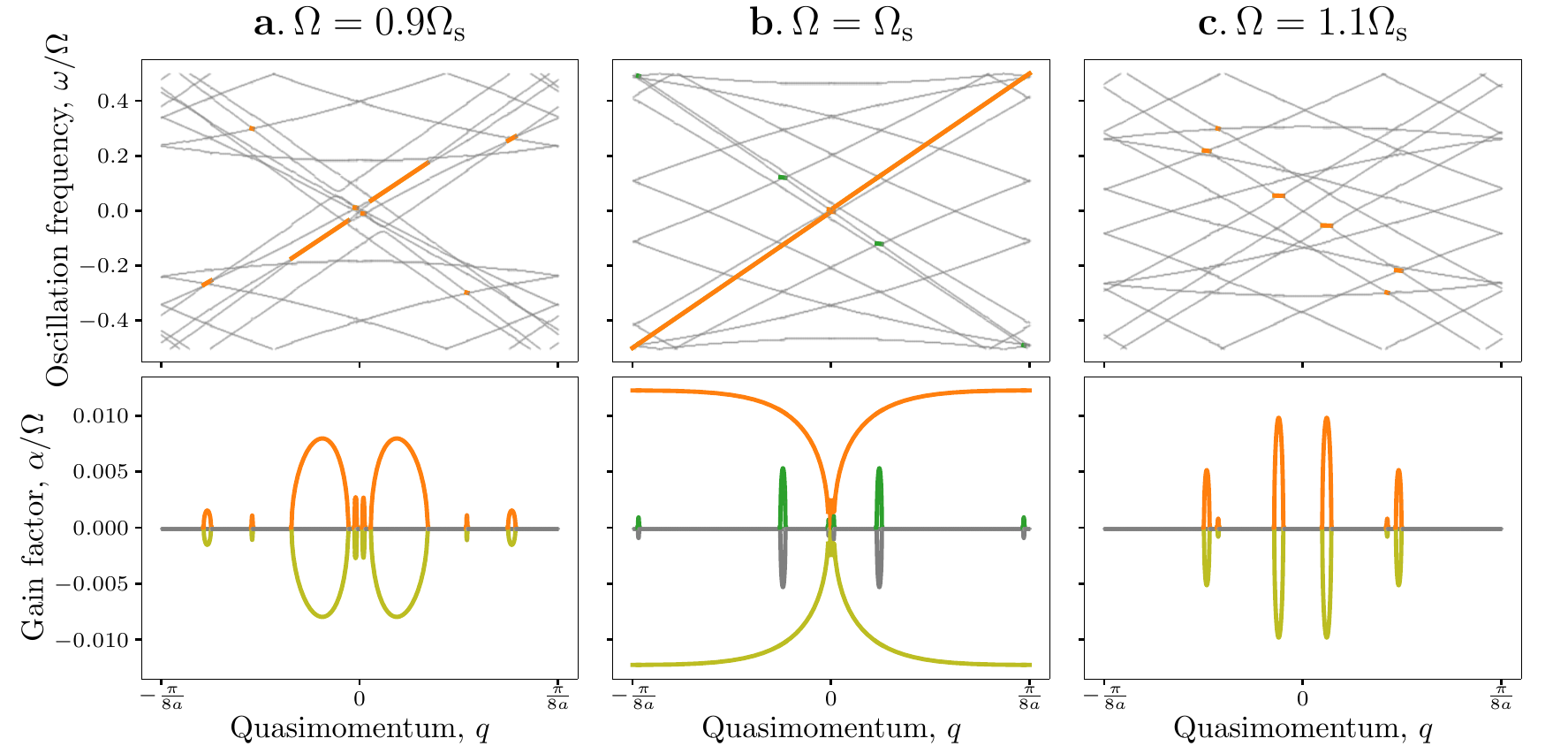}
  \caption{{\bfseries Floquet-Bloch complex band structures near the sonic
      limit.} Real (top) and imaginary (bottom) components of Floquet
    quasifrequencies as a function of crystal quasimomentum for a system with $n=8$
    and a traveling-wave modulation of $\delta = 0.18$, at three different
    modulation frequencies (columns \subfiglabel{a--c}). Floquet-Bloch analysis
    provides $2n=16$ complex-valued bands across the rBZ.
    Thin grey curves show modes with purely
    real frequencies, while thick curves show modes with nonzero gain or loss.
    The latter modes occur in pairs with identical oscillation frequencies,
    one of which is obscured by the other in the top row.
    The middle column shows a system at the sonic limit. }
  \label{fig:bands}
\end{figure*}

As an alternative to considering a system with finite extent, we avoid the
divergences that plague the sonic limit by considering a \emph{discrete} periodic system of
masses $m$ connected by springs, whose vector of displacements $\mathbf{x}$ from
equilibrium is governed by the equation of motion
\begin{equation}
  \label{eq:dyneq-short}
  m \mathbf{\ddot{x}}  + \mathbf{K}(t) \mathbf{x} = 0,
\end{equation}
where $\mathbf{K}$ is the stiffness matrix (see \appref{app:floquet} for the
form of the matrix). 
Such spring-mass lattices comprise a minimal model for vibrational
waves in crystals~\cite{ashcroft2016solid}, and can also be used to describe
effective coupled degrees of freedom in metamaterials with continuum elastic
components~\cite{Matlack2018a,Karki2021}. The normal modes of an infinitely long
periodic chain with lattice spacing $a$ are described as continuous bands over a
restricted set of unique quasimomenta $-\pi/a \leq q < \pi/a$ which defines the
Brillouin zone (BZ). When all masses and springs are constant and equal, the
eigenmodes of the Fourier-transformed
system are organized into two bands with a nonlinear dispersion relation $\omega_\text{s}(q) = \pm 2
\omega_0 \sin ( qa/2)$ (\subfigref{fig:intro}{c}). However, for a
free-standing chain the lowest-frequency, or acoustic, band generically has a
linear dispersion at low quasimomenta because translations of the structure do
not stretch or compress any springs. This translational symmetry
can also be replicated in anchored degrees of freedom coupled by tensile
springs, which do not stretch when all points are displaced by the same amount
(\subfigref{fig:intro}{d}).

We now consider the effect of sinusoidal stiffness modulations which are
themselves periodic in space over a unit cell comprising $n$ degrees of freedom,
\begin{equation}
  \label{eq:travelingwave}
  k_i(t) = k_0\left[1 + \delta \cos \left(\frac{2 \pi}{n}i - \Omega t\right)\right],
\end{equation}
where $k_i$ is the stiffness of the $i$th coupling element along the chain,
$k_0$ and $\delta$ are the base stiffness and the fractional amplitude of the
stiffness modulation respectively, and $\Omega$ is the modulation frequency. The
choice of unit cell defines a range of allowed quasimomenta or reduced Brillouin
zone (rBZ)
of $-\frac{\pi}{na} < k < \frac{\pi}{na}$, shown in grey in \subfigref{fig:intro}{c}.
When $n$ is large, the acoustic band is effectively linear across the entire
rBZ, and the sonic limit corresponds to
\begin{equation}
  \label{eq:sonic}
  \Omega = \frac{2\pi v}{n a} = \frac{2\pi}{n} \omega_0 \equiv \Oms,
\end{equation}
where the speed of sound is dictated by the microscopic parameters via
$v = \sqrt{k_0/m}a = \omega_0 a$ with $\omega_0$ the natural frequency of the
unmodulated springs. In the vicinity of this limit, we expect that modes over a
wide range of frequencies in the lowest frequency band will experience
parametric amplification due to interference with a large number of higher
harmonics similar to \subfigref{fig:intro}{b}.

Since only a finite number of degrees of freedom are involved within each unit
cell, the dispersion relations of the $2n$ bands (arising from the $n$ degrees
of freedom per unit cell for a second-order system of differential equations) of the time-modulated system can
be computed exactly using Floquet-Bloch theory without resorting to coupled-mode
expansions or perturbative treatments, as described in \appref{app:floquet}.
The theory generates complex-valued quasifrequencies
$\nu(q) = \omega(q) + i \alpha(q)$ as a function of quasimomenta $q$ in the
rBZ $-\pi/(na) < q < \pi/(na)$. The real part (which can be
positive or negative) sets the oscillation frequency $\omega$ of the mode,
whereas the imaginary part $\alpha$ signifies exponential growth ($\alpha > 0$)
or decay $(\alpha < 0$) of the underlying mode in time. Floquet theory dictates
that every real-valued mode $\nu(q) = \omega$ is accompanied by a mode at the
opposite quasimomentum with $\nu(-q) = -\omega$. For every mode with complex
quasifrequency $\nu(q)=\nu_0$, a mode at the same quasimomentum with
quasifrequency $\nu(q)=\nu_0^*$ (i.e., same oscillation frequency and gain
factor with same magnitude but opposite sign) is also a solution, as are two
modes at the opposite quasimomentum with $\nu(-q) = -\nu_0$ and
$\nu(-q) = -\nu_0^*$. Furthermore, the real-valued oscillation frequencies are defined modulo
$\Omega$; a minimal set of Floquet-Bloch bands is
therefore defined in the range $-1/2 < \omega/\Omega < 1/2$.

Figure~\ref{fig:bands} shows the Floquet-Bloch bands with lowest oscillation
frequency arising from the acoustic bands of a chain with $n=8$ points in the
unit cell, with traveling-wave modulation frequency below, at, and above the
sonic limit defined by \eqnref{eq:sonic}. As required by the Floquet structure,
modes with complex-valued quasifrequencies occur in pairs with the same
oscillation frequency and opposite gain factors. Away from the sonic limit
(\subfigref{fig:bands}{a,c}) complex bands occur in disconnected segments
separated by quasimomenta with purely real frequencies. Exactly at the sonic
limit $\Omega = \Oms$, the
complex bands with constant positive slope acquire an imaginary component across
the entire rBZ (\subfigref{fig:bands}{b}), with a near-constant value
of the gain factor at large quasimomenta. The effect is directional: whereas the band
with a positive group velocity is amplified throughout,
the accompanying band with a negative slope barely experiences amplification,
except in narrow quasimomentum ranges. These calculations show that unidirectional, broadband amplification of
vibrations can be achieved by modulating spring stiffnesses at the sonic limit.

\section{Strength and parameter dependence of effect} \label{sec:param}

\begin{figure*}
  \centering
  \includegraphics{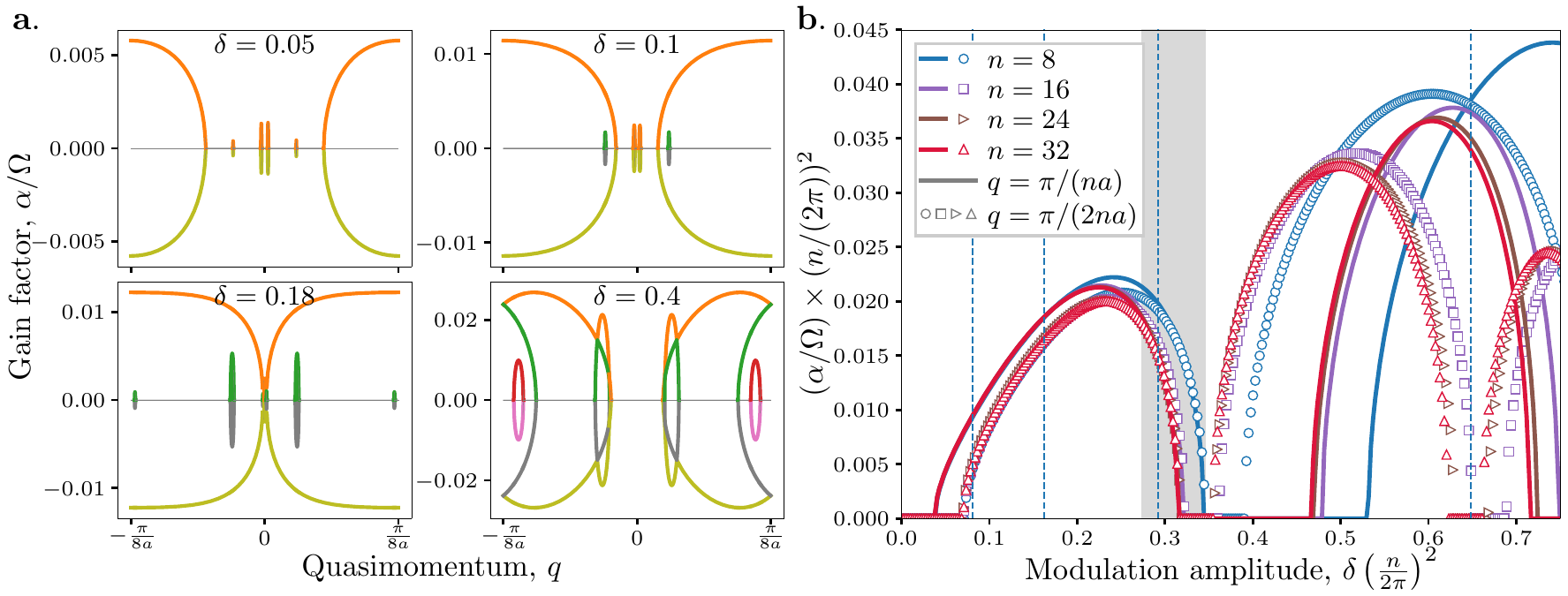}
  \caption{{\bfseries Parameter dependence of gain at the sonic limit.}
    \subfiglabel{a}, Gain factors (imaginary components of Floquet-Bloch bands)
    at $\Omega = \Oms$ for $n=8$ and varying stiffness modulation strength
    $\delta$.
    \subfiglabel{b}, Largest gain factor at the band edge ($q = \pi/(na)$, solid
    lines) and halfway between the origin and the band edge ($q = \pi/(2na)$,
    symbols) as a function of modulation strength for different unit cell sizes
    $n$. Both quantities are scaled by the parameter $(\Oms/\omega_0)^2 =
    (2\pi/n)^2$ which sets the strength of parametric amplification in the
    system. Vertical dashed lines indicate the four values of the modulation
    strength for the $n=8$ bands in \subfiglabel{a}. In the shaded region, the
    gain factors at wavevectors $\pi/(na)$ and $\pi/(2na)$ coincide, consistent
    with broadband amplification of near-constant strength across much of the rBZ.
  }
  \label{fig:param}
\end{figure*}

While a near-constant gain factor can be attained across much of the Brillouin
zone in the sonic limit, this is not guaranteed at all modulation strengths. The
nonlinear dispersion relation of the unmodulated system, and the presence of
additional bands that accidentally satisfy resonance conditions with the traveling-wave
modulation, together generate a rich structure of complex Floquet-Bloch bands at
the sonic limit. Figure~\ref{fig:param}\subfiglabel{a} shows how the gain
factors vary with modulation strength $\delta$ at the sonic limit for a unit
cell with $n=8$. At low modulation strengths, a region of nonzero gain opens
up in the lowest-frequency band with positive slope around the band edges $q =
\pm \pi / (8a)$. The region expands towards the origin, creating a nearly flat
gain-quasimomentum dependence across much of the rBZ at $\delta = 0.18$. 
At higher modulation strengths, several bands
acquire an appreciable gain factor in various quasimomentum ranges due to
additional resonances, and the gain
factors become strongly $q$-dependent. 

Despite this rich structure, at the sonic limit we can reliably find modulation
strengths which realize near-constant broadband amplification. When time is
scaled by the inverse of the modulation frequency $\Omega$, the
Fourier-transformed dynamical matrices depend on the rescaled base stiffness
$(\omega_0/\Omega)^2$ and the rescaled modulation strength
$ \delta (\omega_0/\Omega)^2$. At the sonic limit $\omega_0/\Omega = n/(2\pi)$,
we expect resonances across the lowest band, whose gain factor is set by the
rescaled strength $\delta (n/(2\pi))^2$. Calculations of the largest gain factor
at the BZ edge, $q = \pi/(na)$, and halfway to the BZ center, $q = \pi/(2na)$,
show that the gain factor takes on similar values at both quasimomenta for
rescaled modulation strengths in the range 0.25 to 0.32 before dropping back to
zero. Full Floquet-Bloch band structures for higher values of $n$ confirm
near-constant gain factors across the BZ, see supplementary~\figref{fig:manyn}. At higher
modulation strengths, additional regions of nonzero gain arise, but the gain
factors differ across the band (lower right panel of \subfigref{fig:param}{a}).
These regions are reminiscent of higher-order ``instability tongues'' in the
Mathieu equation, and arise due to additional resonances among the vibrational
modes and the modulation~\cite{Kovacic2018}. The absence of perfect collapse of
the curves at different unit cell sizes is due to the finite deviation of the
lowest band from the idealized linear dispersion, which becomes smaller as $n$
increases due to the shrinking reduced Brillouin zone (\subfigref{fig:intro}{c}).

\subfigureref{fig:param}{b} shows that the fractional stiffness
modulation required for the strongest broadband amplification falls with increasing unit cell
size, $\delta \approx 0.27 (2 \pi/n)^2$. In principle, broadband amplification
can be realized even if the experimentally-achievable stiffness modulation
strength is small, by increasing the wavelength of the stiffness-modulating
traveling wave. However, the corresponding gain factor, which is set by the
modulation strength and governs the
exponential growth $\sim e^{\alpha t}$ in the signal with time, falls as
$\alpha/\Omega \sim \delta \sim 1/n^2$ (or $\alpha/\omega_0 \sim 1/n^3$). At higher
stiffness modulations, the sonic limit extends over a broader range
of modulation phase velocities~\cite{Cassedy1963}, and stronger broadband
amplification can be achieved with slightly slower traveling-wave modulations $\Omega
\lesssim \Oms$ (see \appref{app:largedelta}).

\section{Interplay of parametric amplification and damping}

\begin{figure*}[tb]
  \centering
  \includegraphics{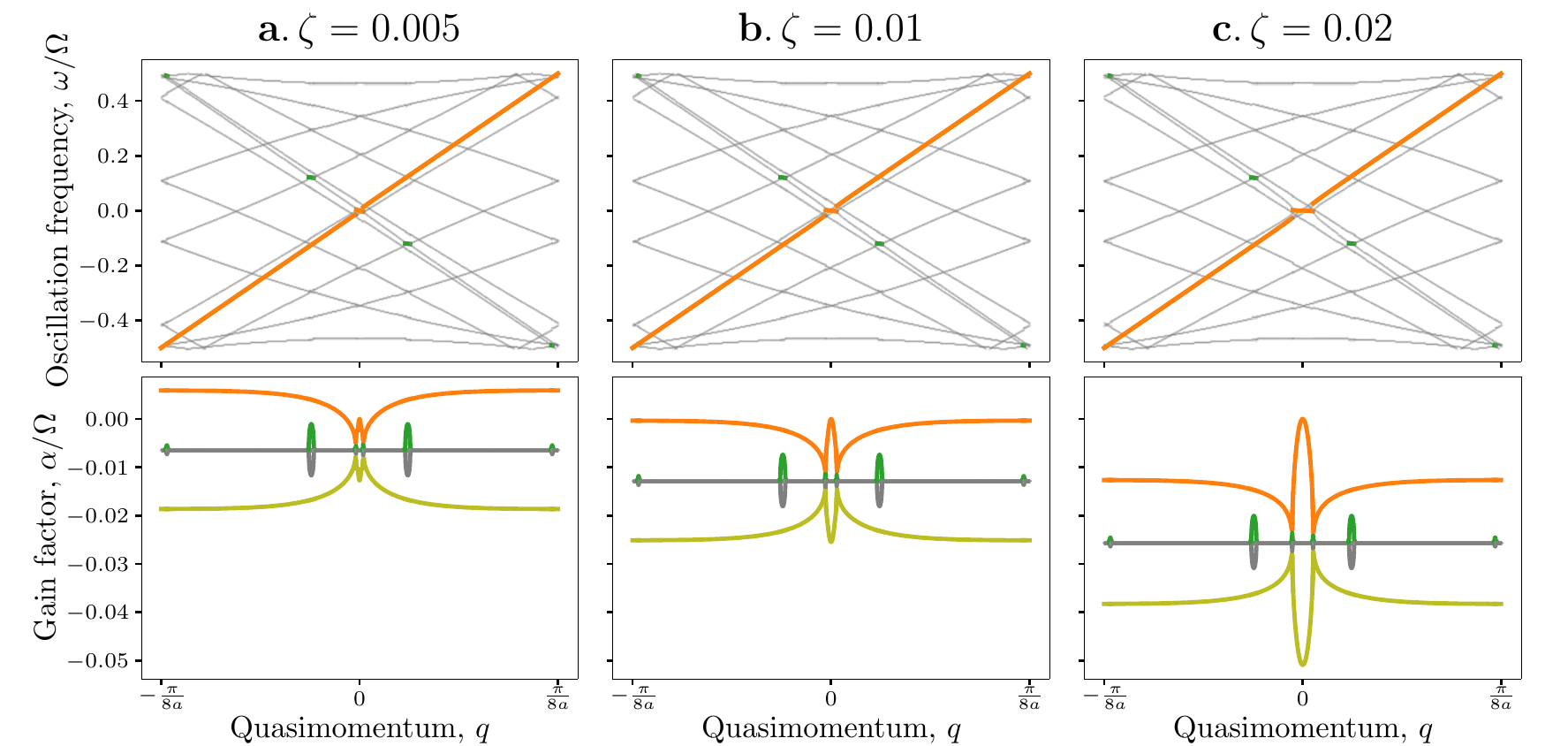}
  \caption{{\bfseries Effect of damping on broadband parametric amplification.}
    Real (top) and imaginary (bottom) components of Floquet 
    quasifrequencies as a function of crystal quasimomentum for a system with
    $n=8$, $\delta = 0.18$, and $\Omega = \Oms$, at three different
    values of the damping ratio $\zeta = \gamma/(2m \omega_0)$. In this case,
    modes that are not parametrically amplified have a negative gain factor
    $\alpha/\Omega = -\zeta \omega/\Omega = -\zeta n / (2\pi)$, signifying exponentially-damped modes, and are shown
    in grey. Thick colored curves show modes whose gain factors deviate from
    this value.
  }
  \label{fig:dampedbands}
\end{figure*}

The existence of modes with positive gain factors points to the presence of
instabilities---even the slightest perturbation to the system, if it overlapped
with one of the amplified modes, would lead to displacements which grow
exponentially in time and ultimately overcome the system. However, if the
amplification can be controlled---for instance, by turning the stiffness
modulation on for finite periods of time---the positive gain factors can be used
to amplify modes in the system. Furthermore, in systems with damping, positive
gain factors can be used to compensate for losses and propagate signals over
longer distances.

We first review the effect of damping on the phonon band structure of a static
spring lattice with uniform spring constants. Upon adding a drag force of
the form $-\gamma \dot{\mathbf{x}}$ to the equation of motion (\eqnref{eq:dyneq-short}), the  frequency bands
become complex-valued:
\begin{equation}
  \label{eq:omegadamped}
  \omega(q) = \pm \sqrt{\omega_\text{s}^2(q) -
  \left(\frac{\gamma}{2m}\right)^2}. 
\end{equation}
Modes whose undamped frequency was large compared to the damping frequency scale
$\gamma/m$ experience a small shift in their oscillatory frequency, and acquire
a negative gain factor $\alpha = -\gamma / (2m)$. At low frequencies, however,
the frequencies become purely imaginary, and the two overdamped modes decay
exponentially with rate $\alpha = -\gamma/(2m) \pm
\sqrt{(\gamma/(2m))^2-\omega_\text{s}^2}$. At $q=0$, a zero-frequency mode is
guaranteed to exist, which corresponds to a displacement of all masses by the
same amount at zero speed, which does not deform any springs and induces no drag
forces. This mode is accompanied by another mode with $\alpha = -\gamma/m$,
corresponding to all masses initially moving at the same speed.

The effect of damping on modulated structures is readily incorporated in the
Floquet-Bloch eigenvalue computation, and leads to similar results to the static
case. \figureref{fig:dampedbands} shows the effect of damping on the sonic
metamaterial with $n=8$ reported in \subfigref{fig:bands}{b}. The damping
strength is quantified by the dimensionless damping factor
$\zeta \equiv \gamma/(2m\omega_0)$. In the presence of drag forces, the
symmetries between complex quasifrequences $\nu(q)$ and $\nu(-q)$ are no longer
obeyed. Instead, we find that modes with large oscillation frequencies compared
to $\gamma/m$ have their gain factors shifted down by roughly $\gamma / (2m)$,
whereas modes near $q = 0$ in the lowest band become purely imaginary. However,
the near-constant value of the gain factor away from $q=0$ is maintained,
showing that the broadband aspect of parametric amplification near the sonic
limit is preserved in damped systems. At $\zeta = 0.01$, the largest gain factor
is close to zero across the entire acoustic band, signifying a balance point
between the broadband parametric amplification and the drag. We will further
investigate this balance, and its consequences for signal propagation, in the
next section.

\section{Dispersion-free amplification and loss mitigation of sound pulses}

\begin{figure*}
  \centering
  \includegraphics{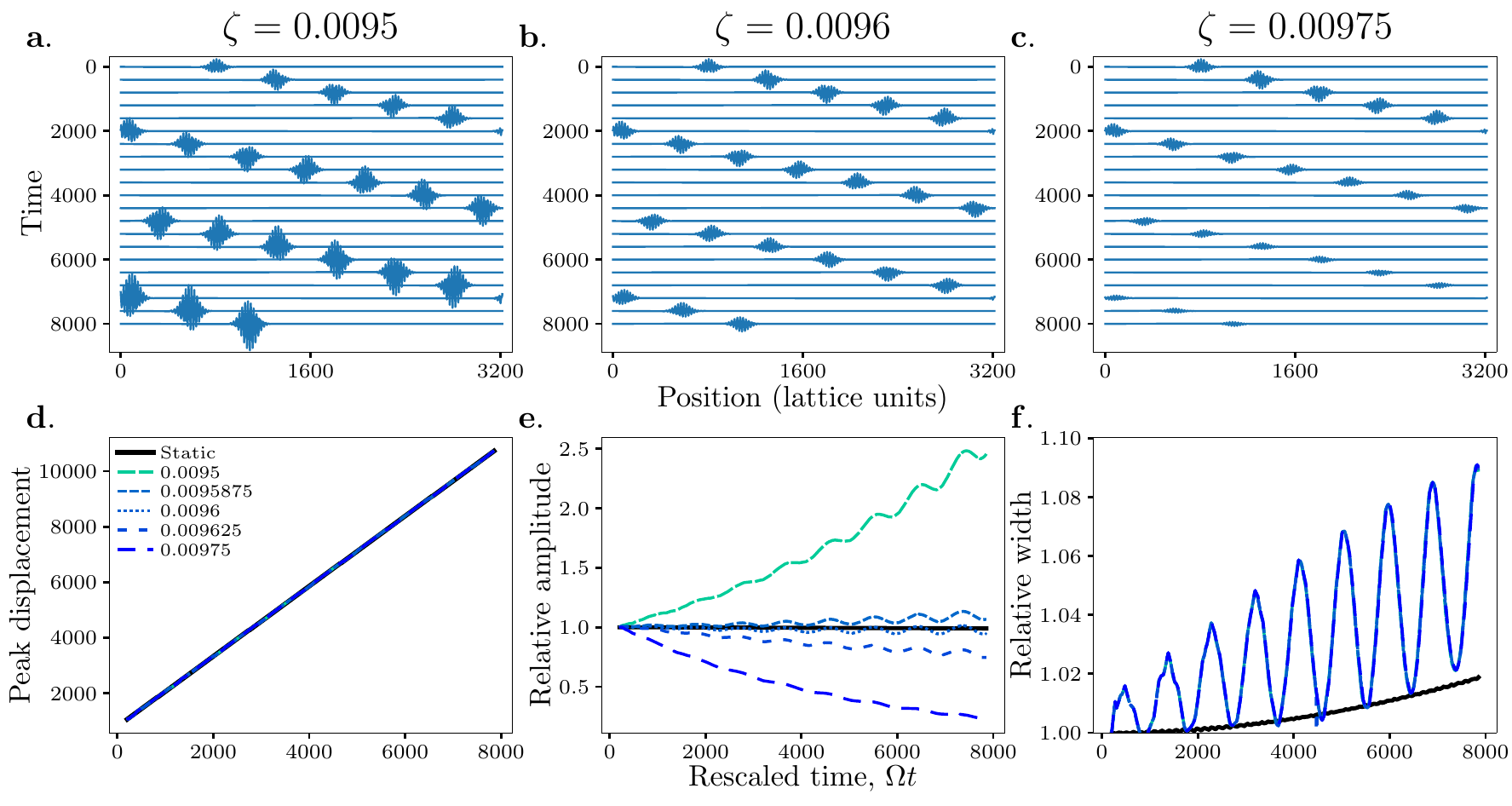}
  \caption{{\bfseries Pulse propagation in classical dynamics simulations of a
      1D spring-mass chain at the sonic limit ($n=8$, $\delta = 0.18$, $\Omega= \Oms$).} %
    \subfiglabel{a--c}, Evolution of the same initial pulse in
    simulations with damping strength above, at, and below the balance point with the
    broadband parametric amplification. Curves are ordered in
    increasing time from top to bottom, and time advances by 10000 timesteps
     between successive curves. Periodic boundary conditions allow the
     propagating pulse to wrap around the system several times.
     \subfiglabel{d--f}, Pulse propagation properties obtained by fitting a
     Gaussian lineshape to the displacement amplitudes, for different damping
     ratios near the balance point.  
    \subfiglabel{d}, Displacement of the pulse peak (in lattice units) as a function of time.
    \subfiglabel{e}, Evolution of the pulse amplitude with time, relative to the
    intial amplitude. Solid line
    corresponds to an undamped system with unmodulated spring stiffnesses
    ($\delta = 0$).
    \subfiglabel{f}, Evolution of the pulse width over time, relative to its
    initial value.
  }
  \label{fig:results}
\end{figure*}

To illustrate the utility of the broadband amplification mechanism for boosting
signals and overcoming losses, we study the propagation of localized sound
pulses (Gaussian wavepackets) along a chain of springs. 
Specifically, the system is initialized with a linear superposition
of eigenmodes $\mathbf{\phi}_q$ from the acoustic band with a linear
frequency-momentum relationship. The mode weights are Gaussian-distributed with spread $\Delta q$ about
the mean quasimomentum $q_0$:
\begin{equation}
  \label{eq:gausspulse}
  \mathbf{u}(n) = \sum_q e^{-[(q-q_0)/\Delta q]^2} \boldsymbol{\phi}_q e^{i(qn - \omega_q t)}, 
\end{equation}
where $\mathbf{u}(n)$ is the vector of initial displacements of the $n$th unit
cell, $\boldsymbol{\phi}_q$ is the Floquet-Bloch eigenvector of the amplified band
at quasimomentum $q$, and $\omega_q$ is the real component of the eigenfrequency. 
In the unmodulated and undamped static system, the resulting sound pulse propagates at a constant
speed given by the slope of the linear dispersion relation, $\partial
\omega/\partial k = v$ (\appref{app:dispersion}).
In the presence of damping, however, all modes decay exponentially in time as
$\sim e^{-\gamma t/(2m)}$, leading to an overall exponential decay in the pulse
amplitude. When the broadband amplification is turned on, the near-constant gain compensates
for the damping across most of the band, shifting the negative gain factors towards or above zero as the
modulation strength is increased. Upon turning on the broadband
modulation at increasing strengths, the pulse attenuation can be slowed down
or even reversed to amplify the pulse as it propagates along
the chain, as shown in~\figref{fig:results}. At a particular modulation
strength, the net gain factors are zero across most of the acoustic band
(\subfigref{fig:dampedbands}{b}), and we expect the sound pulse to travel at
constant amplitude with little dispersion, demonstrating near-ideal loss
compensation through stiffness modulation.

We test this mechanism in classical dynamics simulations of a finite one-dimensional
spring-mass system (see \appref{app:methods} for details) at the sonic limit, with different
damping levels (\figref{fig:results}). The spring constants were modulated
according to the parameters used in \subfigref{fig:bands}{b}. A Gaussian pulse
was initialized with $q_0 = 0.25/a$, $\Delta q = 0.1q_0$ and the subsequent
dynamics of the chain were simulated over thousands of stiffness modulation
cycles. We find that the wavepacket travels at a constant speed, and its
amplitude shows different dynamics depending on the damping strength but with
minimal distortion of the pulse width or shape
(\subfigref{fig:results}{a--c}). In particular, at a specific value of the
damping relative to the modulation, the pulse maintained a near-constant
amplitude over long times as shown in \subfigref{fig:results}{b}.

To assess the fidelity of the loss-compensation
and amplification, we track the pulse position, amplitude, and width as a
function of time by fitting a Gaussian profile to the
displacement field (see Methods for details). Consistent with our theoretical
expectations from the band structure calculations, we find that the wavepacket
speed is not affected by the amplification level (\subfigref{fig:results}{d}).
The relative amplitude (the ratio of the pulse amplitude to its initial value)
shows exponential growth, decay, or stasis as the system damping is changed.
Specifically, a threshold value $\zeta = 0.0096$ separates damping factors at
which the pulse amplitude decreases over time from those for which the amplitude
increases (\subfigref{fig:results}{e}). At this value, damping and parametric
amplification are in balance, signifying idealized loss-compensation in the
system. The pulse width changes by only a few percent over thousands of cycles
of the stiffness modulation (\subfigref{fig:results}{f}). This lack of
dispersion is owed to the nearly constant values of the gain factors across the
band in sonic metamaterials; non-constant gain factors would lead to rapidly changing and
spectral properties, as we show in \appref{app:dispersion}.

\section{Discussion}

We have shown that a traveling-wave stiffness modulation generates broadband
parametric amplification of sound waves when the modulation wave speed
approaches the speed of sound in the medium, confirming a recent hypothesis
grounded in a similar effect for light~\cite{Galiffi2019}. We quantify the
effect using a discrete Floquet-Bloch approach to compute complex-valued
quasifrequencies without encountering divergences or requiring truncated
expansions. For a broad range of parameter values, we find that the
amplification factor is nearly constant over almost all quasimomenta of a
particular band, which enables Gaussian wavepackets to propagate without
dispersion or energy loss even in the presence of damping. The amplification is
highly directional, signifying a strong nonreciprocal response in the
metamaterial~\cite{Nassar2020}. The mechanism could be realized in any active
acoustic metamaterial with a linear dispersion relation at low quasimomenta for
which the effective stiffness can be modulated in space and time, such as beams
with piezoelectric~\cite{Trainiti2019} or electromagnetic~\cite{Chen2019a}
actuation, or backgated micromechanical resonator arrays~\cite{Cha2018}.

Beyond signal amplification, our work suggests several avenues for future research. The presence
of parametric gain in our system makes the underlying eigenvalue problem
non-Hermitian. Our strategy therefore complements approaches based on active
feedback to realize non-Hermitian mechanical
phenomena~\cite{Ghatak2020,Rosa2020,Braghini2021}. The exact Floquet-Bloch
framework used here is equally applicable to slow and fast time modulations,
bridging the gap between theoretical approaches that rely on adiabatic (for slow
modulation) or Magnus (for fast modulation) expansions and thereby enabling the
exploration of non-Hermitian topological phenomena in regimes where the
modulation and excitation frequencies are of similar order~\cite{Coulais2021}.
Higher-dimensional generalizations of the mechanism are also conceivable, since
the nearly-linear dispersion relation at zero quasimomenta is guaranteed by
translational symmetry.

\begin{acknowledgments}
  \acktext
\end{acknowledgments}

\appendix
\setcounter{figure}{0}
\renewcommand{\thefigure}{A\arabic{figure}}

\section{Floquet-Bloch band structures of spacetime-modulated spring lattices} \label{app:floquet}
Prior studies of spring networks with modulated stiffnesses have relied on
various approximations to analyze the eigenmode structure. In the time domain,
the Magnus expansion has been used~\cite{Salerno2016} which relies on a
separation of slow and fast frequency scales in the system. This assumption
breaks down at the sonic limit, where the modulation frequency is comparable to
the normal mode frequencies of the unperturbed system. Alternatively, many
studies use plane-wave expansions of the spatial
eigenmodes~\cite{Zanjani2014,Swinteck2015,Trainiti2016,Nassar2017a,NassarH.2017,Deymier2017,Vila2017,Attarzadeh2018,Attarzadeh2018a,Wang2018,Nassar2018,Li2019a,Chen2019a,Trainiti2019}
which must be truncated at some high wavevector to carry out actual
computations. However, these approaches have been shown to be liable to
inaccuracies at the sonic limit where an ever-larger number of plane waves must
be included in the expansion to avoid divergences in the perturbative
calculations~\cite{Hessel1961,Cassedy1963,Cassedy1967}.

To accurately predict the vibrational modes of spacetime-modulated spring
lattices, we use an exact Floquet-Bloch approach which avoids perturbative expansions, albeit
at the cost of requiring a numerical integration of the underlying dynamical
equations over one time period. Our approach is similar to those used for driven
electronic systems~\cite{Gomez-Leon2013,Holthaus2015},
but adapted to the second-order equations of mechanics~\cite{Iakubovich1975}.
We are interested in the normal modes of a spring-mass chain of $N$ masses, whose displacements
are arranged into an $N$-vector $\mathbf{x}$. When the springs are harmonic, the
equation of motion is
\begin{equation}
  \label{eq:dyneq}
  m \mathbf{\ddot{x}} + \boldsymbol{\Gamma}_N \mathbf{\dot{x}} + \mathbf{K}(t) \mathbf{x} = 0,
\end{equation}
where $\boldsymbol{\Gamma}_N = \gamma \times \mathds{1}_N$ is an $ N\times N$ diagonal matrix of
drag coefficients (assumed uniform), and $\mathbf{K}$ is an $N \times N$ matrix
of spring stiffnesses which encodes the coupling of adjacent degrees of freedom.
For the 1D chain, the stiffness matrix takes the tridiagonal form
\begin{equation}\label{eq:kmatrix}
    \mathbf{K} = 
    \begin{pmatrix}
    ... & ...  & ... & ... & ... & ... & ... & ... \\
    ... & 0 & -k_j & k_j + k_{j+1} & - k_{j+1} & 0 & ... & ... \\
    ... & 0 & 0 & -k_{j+1} & k_{j+1} + k_{j+2} & - k_{j+2} & 0 & ... \\
    ... & ...  & ... & ... & ... & ... & ... & ... 
  \end{pmatrix}
\end{equation}
When the spring constants $k_j$ are modulated in time and space according to the
traveling-wave modulation
\begin{equation}
  \label{eq:travelingwave-supp}
  k_j(t) = k_0\left[1 + \delta \cos \left(\frac{2 \pi}{n}j - \Omega t\right)\right],
\end{equation}
the eigenmodes of the dynamical system can be
written in terms of an $n$-vector $\mathbf{u}_q(t)$ and a quasimomentum $q$, where
the displacements of the $p$th unit cell at position $x = pna$ are given by
$\mathbf{u}_q(t) e^{i q x}$. The $\mathbf{u}_q(t)$ solve the equation  
\begin{equation}
  \label{eq:dyneq-q}
  m \mathbf{\ddot{u}}_q + \boldsymbol{\Gamma}_n \mathbf{\dot{u}}_q + \mathbf{\tilde{K}}(q,t) \mathbf{u}_q = 0,
\end{equation}
where the Fourier-transformed stiffness matrix $\mathbf{\tilde{K}}(q,t)$ has dimensions $n \times n$, and
includes phase factors $e^{\pm i q na}$ for springs that extend to neighboring
unit cells. For an infinite periodic lattice, the periodicity defines a unique
set of quasimomenta $-\pi < q na \leq \pi$, which define the reduced Brillouin zone.

We now exploit the time-periodicity of the stiffness matrix, $\mathbf{\tilde{K}}(q,t+T)=\mathbf{\tilde{K}}(q,t)$ where $T=2\pi/\Omega$. To apply the Floquet theory of first-order
differential equations, we rewrite
\eqnref{eq:dyneq-q} as a first-order equation involving the doubled vector
$\mathbf{y}_q = (\mathbf{u}_q, \mathbf{\dot{u}}_q)^\top$, 
\begin{equation}
  \label{eq:floquet-1}
  \mathbf{\dot{y}}_q = \mathbf{G}_q(t) \mathbf{y}_q,
\end{equation}
where $$\mathbf{G}_q(t) = \begin{pmatrix} 0 & \mathds{1}_n \\ -\mathbf{\tilde{K}}(q,t) & -\mathbf{\Gamma}_n \end{pmatrix}$$ 
inherits the time-periodicity of the stiffness matrix. Any solution to the
differential equation can be written in terms of the matrix of solutions, 
$\mathbf{X}(t)$, which satisfies
\begin{equation}
  \label{eq:matrizant}
  \mathbf{\dot{X}} = \mathbf{G}_q(t) \mathbf{X},
\end{equation}
starting from the initial condition $\mathbf{X}(0) = \mathds{1}_{2n}$.
Any solution of \eqnref{eq:floquet-1} then can be
written in terms of the initial condition as
$$\mathbf{x}(t) = \mathbf{X}(t) \mathbf{x}(0).$$

When $G$ is $T$-periodic, the matrix of solutions has the property
\begin{equation}
  \label{eq:floquet-2}
  \mathbf{X}(t+T) = \mathbf{X}(t)\mathbf{X}(T).
\end{equation}
The solution matrix evaluated over one period, $\mathbf{X}(T)$, is called the
\emph{monodromy matrix}. The eigenvalues $\rho_j$ (with $j = 1,...,2n$) and
corresponding eigenvectors $\mathbf{a}_j$ of the monodromy matrix have the
following useful property: a solution $\mathbf{x}_j(t)$ of \eqnref{eq:floquet-1}
with initial value $\mathbf{x}_j(0) = \mathbf{a}_j$ satisfies
\begin{equation}
  \label{eq:floquet-3}
  \mathbf{x}_j(t+T) = \rho_j \mathbf{x}_i(t).
\end{equation}
The eigenvalues $\rho_j$ are called the \emph{Floquet multipliers} of the
system. 
\equationref{eq:floquet-3} implies the form
\begin{equation}
  \label{eq:quasiperiodic}
  \mathbf{x}_j(t) = e^{-i \nu_j t}\mathbf{f}_j(t),
\end{equation}
where $\nu_j \equiv i(\ln \rho_j)/T$ is the $i$th \emph{Floquet quasifrequency}, and
$$\mathbf{f}_j(t) = \mathbf{X}(t) e^{-\frac{t}{T}\ln \mathbf{X}(T)}\mathbf{a}_j$$ is
$T$-periodic by the periodicity of the matrix of solutions:
$$\mathbf{f}_j(t+T) = \mathbf{f}_j(t).$$ The $2n$ vectors $\mathbf{x}_j(t)$ are
linearly independent and form a fundamental set of solutions of the system. 

At each quasimomentum $q$, the Floquet calculation gives us $2n$ quasifrequencies $\nu_j(q) = \omega_j(q) + i \alpha_j(q)$; these are the Floquet-Bloch bands of
the system. The calculation involves a numerical integration of
\eqnref{eq:matrizant} over one time period. Provided the numerical integration
can be carried out to the desired precision, the calculation of the bands is
exact as it does not rely on any truncated expansion of the solutions in terms
of plane waves.  The corresponding normal mode displacements and velocities of the
$p$th unit cell at position $x = pna$ are written in terms of the Floquet-Bloch eigenvectors
$\mathbf{f}_j(q,t)$ as
\begin{equation}
  \label{eq:floquetblochmode}
  \mathbf{y}_j(q,t) = \mathbf{f}_j(q,t)e^{i[q x - \nu_j(q)t]}.
\end{equation}
This form is similar to that of the normal modes of a \emph{static} spring
chain, with the differences that: i. the vector multiplying the plane-wave itself has an
additional time dependence (albeit one that is $T$-periodic in time); ii. the Floquet
exponents which take the place of the frequencies are in general complex-valued;
iii. the oscillation frequencies $\omega_j = \operatorname{Re}(\nu_j)$ are defined modulo the
modulation frequency $\Omega$. 

The correspondence of the Floquet-Bloch eigenmodes with normal modes of static
systems is even stronger if we consider \emph{strobed} measurements, i.e. when
displacements are recorded only at integer multiples of the time period $T$.
At these time intervals, we have $$\mathbf{y}_j(q,t) = \mathbf{f}_j(q,0)e^{i[qx -
  \nu_j(t)]};$$ i.e. the strobed spacetime-dependence is obtained by multiplying
a constant eigenvector with a plane wave. When
measurements are strobed, therefore, the Floquet eigenvectors and exponents are
completely analogous to the normal modes and eigenfrequencies of a static spring
network. The group and phase velocities of waves in the $j$th band are
determined by the dispersion relation $\omega_j(q)$. When the gain factor is
nonzero, a positive gain factor corresponds to an exponentially growing wave
amplitude $\propto e^{\alpha_j t}$ whereas a negative gain factor corresponds
to an exponentially decaying wave.

\section{Sonic limit at larger modulation strengths} \label{app:largedelta}

\begin{figure*}[t]
  \centering
  \includegraphics{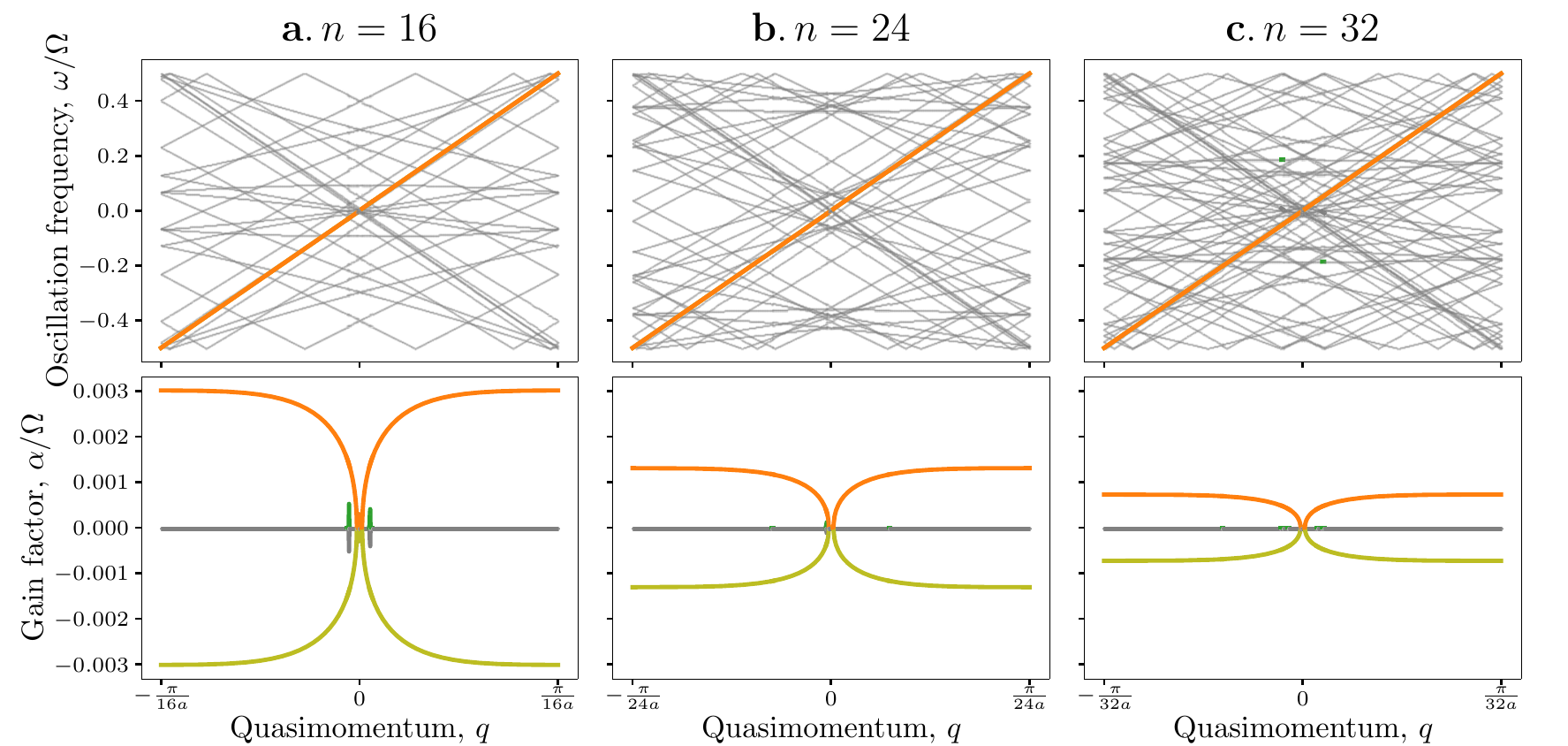}
  \caption{{\bfseries Floquet-Bloch band structures with $\Omega = \Oms$ and
      $\delta = 0.27 (2\pi/n)^2$ for different unit cell sizes.} While the
    number of bands $2n$ grows with unit cell size, only the acoustic band
    with positive slope is singled out for broadband parametric amplification,
    with an associated gain factor that scales as $1/n^2$. }
  \label{fig:manyn}
\end{figure*}

\begin{figure}
  \centering
  \includegraphics{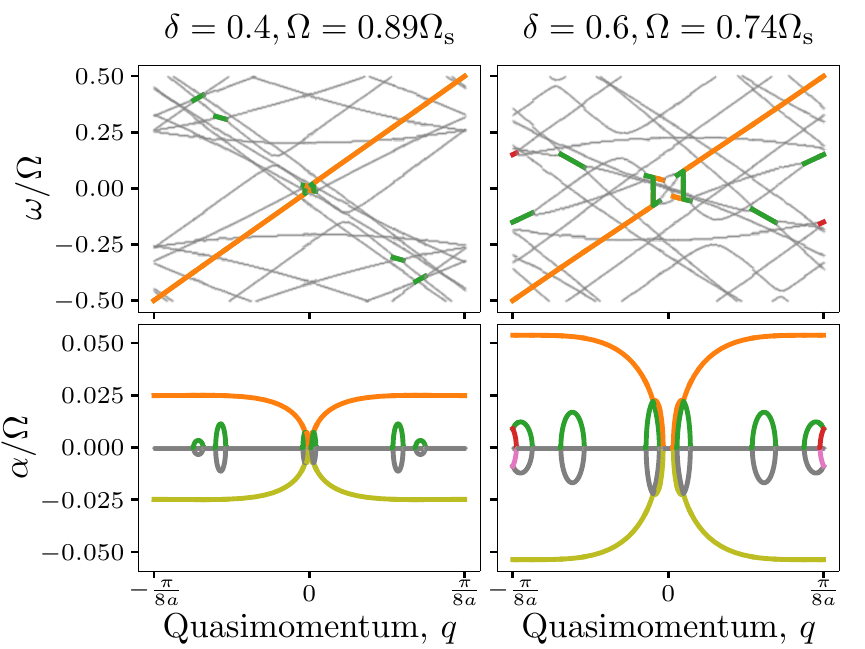}
  \caption{{\bfseries Broadband amplification at large modulation strengths.}
    Floquet-Bloch band structures, similar to those in the main text, for
    systems with $n=8$ and two larger values of $\delta$. The sonic limit
    extends over a range of values around $\Omega = \Oms$ for large modulation
    strengths, and nearly constant gain factors are obtained at lower values of
    $\Omega$ compared to the limit of small $\delta$.
  }
  \label{fig:largedelta}
\end{figure}

The accumulation of resonances which defines the sonic limit extends over a
finite range of modulation frequencies on either side of the value
$\Oms = 2\pi v/(na)$, defined by~\cite{Hessel1961,Cassedy1963}
\begin{equation}
  \label{eq:sonicfull}
  \frac{1}{\sqrt{1+\delta}} < \frac{\Omega}{\Oms} < \frac{1}{\sqrt{1-\delta}}.
\end{equation}
In the main text, we focused on modulation frequencies at the center of this
range, which is appropriate when $\delta \ll 1$.
However, if larger modulation strengths are accessible, the modulation frequency
that generates near-uniform gain factors across the reduced Brillouin zone can
deviate from the strict sonic limit defined in \eqnref{eq:sonic} of the main
text. In this case, modulation parameters which allow stronger
broadband amplification can be found by numerically exploring Floquet-Bloch
band structures within the range of values dictated by \eqnref{eq:sonicfull}. As
an example, \figref{fig:largedelta} shows near-constant gain factors across the
rBZ for $\Omega = 0.89 \Oms$ when $\delta = 0.4$ and $\Omega = 0.74 \Oms$ when
$\delta = 0.6$, for the same unit cell size ($n=8$) discussed in
\subfigref{fig:param}{a}. In both cases, the larger modulation also enables
higher gain factors to be realized compared to the case of $\Omega = \Oms$.

\section{Simulation methods} \label{app:methods}

Wave pulse propagation in 1D spring-mass chains was studied using classical
dynamic simulations implemented in the
HOOMD-Blue software package~\cite{anderson2008,glaser2015}. The system examined
was a 1D spring-mass chain of $N$ particle-masses of mass $m$ possessing a unit
cell of $n$ different dynamic springs with equilibrium length $l_0$ in a damped
environment with damping constant $\gamma$. To implement dynamic springs
following \eqnref{eq:travelingwave}, the spring constant of each spring was
updated after each time step. 
Simulations were initialized with a Gaussian wavepacket assembled using the
Floquet-Bloch eigenvectors of the chain as described in the text
(\eqnref{eq:gausspulse}).
The Floquet-Bloch calculation generates both positions and velocities
(\appref{app:floquet}) for the initial condition.

For all simulations, we set $m = 1$, $l_0 = 1$, and $k_0=1$ in simulation units.
The spring-mass chain was created using 201 repetitions of a unit cell of $n=8$,
giving rise to a system size of $N=1608$,
with periodic boundary conditions along the $x$ direction. Simulations were
run for $1\times10^8$ time-steps with a step size $\Delta t = 0.0001$. We
checked that reducing the step size by a factor of 4 did not
significantly change the pulse evolution with time.

The overall shape of the wavepacket was tracked during the simulation by fitting
the particle displacement amplitudes to a Gaussian profile for each time
snapshot. The center position, amplitude, and standard deviation of the Gaussian
profile were treated as free parameters whose best-fit values were obtained
using the \texttt{optimize.curve\_fit} function from the \texttt{scipy} package
in Python. These quantities are reported in \figref{fig:results} of the main text.

\section{Dispersion of wavepackets under non-constant gain
  factors} \label{app:dispersion}

Here, we compute the distortion of Gaussian wavepackets whose constituent modes
have non-uniform gain factors. For simplicity, we work with continuum plane
waves of the form $e^{i[k x - \nu(k)t]}$ with momentum $k$ and a 
dispersion relation $\nu(k) = \omega(k) + i \alpha(k)$ relating the complex
frequencies to the wavevectors. We expect the
behavior of the amplitude envelope to be similar for a wavepacket built from
Floquet-Bloch eigenmodes of the amplified band in the discrete system with
time-modulated springs.

First, we review the effect of a nonlinear dispersion relation $\omega(k)$ on the time dynamics of a
Gaussian wavepacket in the absence of gain, $\alpha(k)= 0$. The wavepacket is a
superposition of plane waves with weights
\begin{equation}
  \label{eq:wpinit-k}
  \tilde{f}(k) = A_0 \frac{\sigma}{\sqrt{2\pi}} e^{-\frac{\sigma^2}{2}(k-k_0)^2},
\end{equation}
which leads to a real-space pulse at time $t=0$ of
\begin{equation}
  \label{eq:wpinit-x}
  f(x,0) = A_0 e^{ik_0 x} e^{-\frac{x^2}{2 \sigma^2}},
\end{equation}
where $A_0$ is the initial amplitude, and $\sigma$ is the width of the Gaussian envelope centered at $x=0$ of a
sinusoidally varying wave with dominant wavevector $k_0$. The subsequent
time-evolution is given by
$$ f(x,t) = \int_{-\infty}^\infty \!dk e^{i[kx-\omega(k)t]} \tilde{f}(k).$$
If the pulse is sufficiently broad, the Fourier amplitudes fall off fast away
from $k_0$, and we can approximate the dispersion relation near $k_0$ as a
Taylor series:
$$\omega(k) \approx v_\text{p} k_0 + v_\text{g}(k-k_0) +
\frac{\eta}{2}(k-k_0)^2,$$ where $v_\text{p} = \omega(k_0)/k_0$ and $v_\text{g}
= \partial \omega(k_0)/\partial k$ are respectively the phase and group velocity
of the wave at $k_0$. The solution in real space at finite times is then
obtained by taking the inverse Fourier transform, with the result
\begin{equation}
  \label{eq:wavepackettime}
  f(x,t) = A_0 e^{i(v_\text{g}-v_\text{p})k_0t} e^{i k_0 (x-v_\text{g}t)}
e^{-\frac{(x-v_\text{g}t)^2}{2(\sigma^2 - i \eta t)}}.
\end{equation}

When the dispersion relation is strictly linear, $\eta = 0$, the finite-time
solution has the form $f(x,t) = e^{i(v_\text{g}-v_\text{p})k_0}
f(x-v_\text{g}t,0)$, which corresponds to a translation of the Gaussian
amplitude profile by $v_\text{g}t$ along the $x$-axis, and an additional phase
factor which does not affect the amplitude (and which is zero for a linear
dispersion relation $\omega = v_\text{g} k = v_\text{p} k$). The wavepacket is
said to be non-dispersive, as it maintains its shape while propagating at a
constant speed.

By contrast, when $\eta \neq 0$, the last exponential in
\eqnref{eq:wavepackettime} can be written as
$$e^\frac{i\eta t}{2(\sigma^4+\eta^2 t^2)} e^{-\frac{(x-v_\text{g}t)^2}{2\sigma^2(1+\eta^2 t^2/\sigma^4)}}.$$
Besides introducing an additional phase, the nonzero quadratic dispersion also
modifies the Gaussian amplitude profile which, while still moving with the group
velocity, is rapidly broadening with time as $\sigma\sqrt{1+\eta^2
  t^2/\sigma^4}$. 
Such a wavepacket whose amplitude and phase profile are
varying with time is termed
\emph{dispersive}. Deviations from a linear dispersion involving higher powers
of $k-k_0$ also lead to dispersive pulse propagation. A slight dispersion is
apparent in the simulated time-evolution of pulse width in a
static system (\subfigref{fig:results}{f}), which grows quadratically in time
because of the deviation of the dispersion relation $\omega_\text{s}(q)$ from linearity.

We now consider wavepacket dispersion due to non-zero gain. We consider a linear
dispersion relation of the oscillatory frequency, but assume the gain factor has
a linear wavevector dependence near $k_0$,
\begin{equation}
  \label{eq:complexdisprelation}
  \omega(k) \approx vk + i \left[\alpha_0 + \beta(k-k_0)\right].
\end{equation}
Upon initializing the wavepacket using \eqnref{eq:wpinit-k} and computing the inverse
Fourier transform, we find the subsequent time evolution
\begin{equation}
  \label{eq:wp-amp}
  f(x,t) = A_0  e^{i \left(k_0 + \frac{\beta t}{\sigma^2}\right) (x-vt)}
e^{-\frac{(x-vt)^2}{2\sigma^2}} e^{\alpha_0 t + \frac{\beta^2 t^2}{2\sigma^2}}.
\end{equation}
When $\beta = 0$, a constant gain factor induces an exponential growth of the
wavepacket amplitude with time, but the relative strengths and phases of various
components of the wavepacket are unchanged. This situation corresponds to a
non-dispersive amplification of the overall wavepacket as it propagates. By
contrast, a nonzero slope to the gain-wavevector relation changes the dominant
(or carrier) wavevector of the signal, which increases linearly with time as
$k_0 + \beta t/\sigma^2$. The pulse envelope is still centered at $x-vt$ and
grows with time, but with a additional time-dependence which grows as $\exp(t^2)$, much faster
than exponentially with time. This superamplification arises from the modes with
$k \gg k_0$ when $\beta >0$ (or $k \ll k_0$ for $\beta < 0$) which fall outside
the regime of validity of the candidate dispersion relation,
\eqnref{eq:complexdisprelation}. In practice, the shape and speed of the
wavepacket will depend on the full dispersion relation $\omega(k)$ as the
dominant wavevectors in the wavepacket are no longer confined to a small range
near $k_0$, leading to ever-increasing distortion of the wavepacket.

In summary, Gaussian wavepackets can be amplified without affecting their
spectral composition provided the gain factor is constant over the entire range of
wavevectors contributing to the wavepacket.


\begin{thebibliography}{50}%
\makeatletter
\providecommand \@ifxundefined [1]{%
 \@ifx{#1\undefined}
}%
\providecommand \@ifnum [1]{%
 \ifnum #1\expandafter \@firstoftwo
 \else \expandafter \@secondoftwo
 \fi
}%
\providecommand \@ifx [1]{%
 \ifx #1\expandafter \@firstoftwo
 \else \expandafter \@secondoftwo
 \fi
}%
\providecommand \natexlab [1]{#1}%
\providecommand \enquote  [1]{``#1''}%
\providecommand \bibnamefont  [1]{#1}%
\providecommand \bibfnamefont [1]{#1}%
\providecommand \citenamefont [1]{#1}%
\providecommand \href@noop [0]{\@secondoftwo}%
\providecommand \href [0]{\begingroup \@sanitize@url \@href}%
\providecommand \@href[1]{\@@startlink{#1}\@@href}%
\providecommand \@@href[1]{\endgroup#1\@@endlink}%
\providecommand \@sanitize@url [0]{\catcode `\\12\catcode `\$12\catcode
  `\&12\catcode `\#12\catcode `\^12\catcode `\_12\catcode `\%12\relax}%
\providecommand \@@startlink[1]{}%
\providecommand \@@endlink[0]{}%
\providecommand \url  [0]{\begingroup\@sanitize@url \@url }%
\providecommand \@url [1]{\endgroup\@href {#1}{\urlprefix }}%
\providecommand \urlprefix  [0]{URL }%
\providecommand \Eprint [0]{\href }%
\providecommand \doibase [0]{http://dx.doi.org/}%
\providecommand \selectlanguage [0]{\@gobble}%
\providecommand \bibinfo  [0]{\@secondoftwo}%
\providecommand \bibfield  [0]{\@secondoftwo}%
\providecommand \translation [1]{[#1]}%
\providecommand \BibitemOpen [0]{}%
\providecommand \bibitemStop [0]{}%
\providecommand \bibitemNoStop [0]{.\EOS\space}%
\providecommand \EOS [0]{\spacefactor3000\relax}%
\providecommand \BibitemShut  [1]{\csname bibitem#1\endcsname}%
\let\auto@bib@innerbib\@empty
\bibitem [{\citenamefont {Cullen}(1958)}]{Cullen1958}%
  \BibitemOpen
  \bibfield  {author} {\bibinfo {author} {\bibfnamefont {A.~L.}\ \bibnamefont
  {Cullen}},\ }\href {\doibase 10.1038/181332a0} {\bibfield  {journal}
  {\bibinfo  {journal} {Nature}\ }\textbf {\bibinfo {volume} {181}},\ \bibinfo
  {pages} {332} (\bibinfo {year} {1958})}\BibitemShut {NoStop}%
\bibitem [{\citenamefont {Tien}(1958)}]{Tien1958}%
  \BibitemOpen
  \bibfield  {author} {\bibinfo {author} {\bibfnamefont {P.~K.}\ \bibnamefont
  {Tien}},\ }\href {\doibase 10.1063/1.1723440} {\bibfield  {journal} {\bibinfo
   {journal} {Journal of Applied Physics}\ }\textbf {\bibinfo {volume} {29}},\
  \bibinfo {pages} {1347} (\bibinfo {year} {1958})}\BibitemShut {NoStop}%
\bibitem [{\citenamefont {Baumgartner}\ and\ \citenamefont
  {Byer}(1979)}]{Baumgartner1979}%
  \BibitemOpen
  \bibfield  {author} {\bibinfo {author} {\bibfnamefont {R.}~\bibnamefont
  {Baumgartner}}\ and\ \bibinfo {author} {\bibfnamefont {R.}~\bibnamefont
  {Byer}},\ }\href@noop {} {\bibfield  {journal} {\bibinfo  {journal} {IEEE
  Journal of Quantum Electronics}\ }\textbf {\bibinfo {volume} {15}},\ \bibinfo
  {pages} {432} (\bibinfo {year} {1979})}\BibitemShut {NoStop}%
\bibitem [{\citenamefont {Rugar}\ and\ \citenamefont
  {Gr{\"u}tter}(1991)}]{Rugar1991}%
  \BibitemOpen
  \bibfield  {author} {\bibinfo {author} {\bibfnamefont {D.}~\bibnamefont
  {Rugar}}\ and\ \bibinfo {author} {\bibfnamefont {P.}~\bibnamefont
  {Gr{\"u}tter}},\ }\href {\doibase 10.1103/PhysRevLett.67.699} {\bibfield
  {journal} {\bibinfo  {journal} {Physical Review Letters}\ }\textbf {\bibinfo
  {volume} {67}},\ \bibinfo {pages} {699} (\bibinfo {year} {1991})}\BibitemShut
  {NoStop}%
\bibitem [{\citenamefont {Landau}\ and\ \citenamefont
  {Lifshitz}(1982)}]{landau1982mechanics}%
  \BibitemOpen
  \bibfield  {author} {\bibinfo {author} {\bibfnamefont {L.}~\bibnamefont
  {Landau}}\ and\ \bibinfo {author} {\bibfnamefont {E.}~\bibnamefont
  {Lifshitz}},\ }\href@noop {} {\emph {\bibinfo {title} {Mechanics: {{Volume}}
  1}}},\ \bibinfo {number} {v. 1}\ (\bibinfo  {publisher} {{Elsevier
  Science}},\ \bibinfo {year} {1982})\BibitemShut {NoStop}%
\bibitem [{\citenamefont {Cassedy}\ and\ \citenamefont
  {Oliner}(1963)}]{Cassedy1963}%
  \BibitemOpen
  \bibfield  {author} {\bibinfo {author} {\bibfnamefont {E.~S.}\ \bibnamefont
  {Cassedy}}\ and\ \bibinfo {author} {\bibfnamefont {A.~A.}\ \bibnamefont
  {Oliner}},\ }\href {\doibase 10.1109/PROC.1963.2566} {\bibfield  {journal}
  {\bibinfo  {journal} {Proceedings of the IEEE}\ }\textbf {\bibinfo {volume}
  {51}},\ \bibinfo {pages} {1342} (\bibinfo {year} {1963})}\BibitemShut
  {NoStop}%
\bibitem [{\citenamefont {Cassedy}(1967)}]{Cassedy1967}%
  \BibitemOpen
  \bibfield  {author} {\bibinfo {author} {\bibfnamefont {E.~S.}\ \bibnamefont
  {Cassedy}},\ }\href {\doibase 10.1109/PROC.1967.5775} {\bibfield  {journal}
  {\bibinfo  {journal} {Proceedings of the IEEE}\ }\textbf {\bibinfo {volume}
  {55}},\ \bibinfo {pages} {1154} (\bibinfo {year} {1967})}\BibitemShut
  {NoStop}%
\bibitem [{\citenamefont {Boardman}\ \emph {et~al.}(2011)\citenamefont
  {Boardman}, \citenamefont {Grimalsky}, \citenamefont {Kivshar}, \citenamefont
  {Koshevaya}, \citenamefont {Lapine}, \citenamefont {Litchinitser},
  \citenamefont {Malnev}, \citenamefont {Noginov}, \citenamefont {Rapoport},\
  and\ \citenamefont {Shalaev}}]{Boardman2011}%
  \BibitemOpen
  \bibfield  {author} {\bibinfo {author} {\bibfnamefont {A.~D.}\ \bibnamefont
  {Boardman}}, \bibinfo {author} {\bibfnamefont {V.~V.}\ \bibnamefont
  {Grimalsky}}, \bibinfo {author} {\bibfnamefont {Y.~S.}\ \bibnamefont
  {Kivshar}}, \bibinfo {author} {\bibfnamefont {S.~V.}\ \bibnamefont
  {Koshevaya}}, \bibinfo {author} {\bibfnamefont {M.}~\bibnamefont {Lapine}},
  \bibinfo {author} {\bibfnamefont {N.~M.}\ \bibnamefont {Litchinitser}},
  \bibinfo {author} {\bibfnamefont {V.~N.}\ \bibnamefont {Malnev}}, \bibinfo
  {author} {\bibfnamefont {M.}~\bibnamefont {Noginov}}, \bibinfo {author}
  {\bibfnamefont {Y.~G.}\ \bibnamefont {Rapoport}}, \ and\ \bibinfo {author}
  {\bibfnamefont {V.~M.}\ \bibnamefont {Shalaev}},\ }\href@noop {} {\bibfield
  {journal} {\bibinfo  {journal} {Laser \& Photonics Reviews}\ }\textbf
  {\bibinfo {volume} {5}},\ \bibinfo {pages} {287} (\bibinfo {year}
  {2011})}\BibitemShut {NoStop}%
\bibitem [{\citenamefont {{Zangeneh-Nejad}}\ and\ \citenamefont
  {Fleury}(2019)}]{Zangeneh-Nejad2019}%
  \BibitemOpen
  \bibfield  {author} {\bibinfo {author} {\bibfnamefont {F.}~\bibnamefont
  {{Zangeneh-Nejad}}}\ and\ \bibinfo {author} {\bibfnamefont {R.}~\bibnamefont
  {Fleury}},\ }\href {\doibase 10.1016/j.revip.2019.100031} {\bibfield
  {journal} {\bibinfo  {journal} {Reviews in Physics}\ }\textbf {\bibinfo
  {volume} {4}},\ \bibinfo {pages} {100031} (\bibinfo {year}
  {2019})}\BibitemShut {NoStop}%
\bibitem [{\citenamefont {Wang}\ \emph {et~al.}(2020)\citenamefont {Wang},
  \citenamefont {Wang}, \citenamefont {Wu}, \citenamefont {Chen},\ and\
  \citenamefont {Wang}}]{Wang2020}%
  \BibitemOpen
  \bibfield  {author} {\bibinfo {author} {\bibfnamefont {Y.-F.}\ \bibnamefont
  {Wang}}, \bibinfo {author} {\bibfnamefont {Y.-Z.}\ \bibnamefont {Wang}},
  \bibinfo {author} {\bibfnamefont {B.}~\bibnamefont {Wu}}, \bibinfo {author}
  {\bibfnamefont {W.}~\bibnamefont {Chen}}, \ and\ \bibinfo {author}
  {\bibfnamefont {Y.-S.}\ \bibnamefont {Wang}},\ }\href {\doibase
  10.1115/1.4046222} {\bibfield  {journal} {\bibinfo  {journal} {Applied
  Mechanics Reviews}\ }\textbf {\bibinfo {volume} {72}} (\bibinfo {year}
  {2020}),\ 10.1115/1.4046222}\BibitemShut {NoStop}%
\bibitem [{\citenamefont {Galiffi}\ \emph {et~al.}(2019)\citenamefont
  {Galiffi}, \citenamefont {Huidobro},\ and\ \citenamefont
  {Pendry}}]{Galiffi2019}%
  \BibitemOpen
  \bibfield  {author} {\bibinfo {author} {\bibfnamefont {E.}~\bibnamefont
  {Galiffi}}, \bibinfo {author} {\bibfnamefont {P.~A.}\ \bibnamefont
  {Huidobro}}, \ and\ \bibinfo {author} {\bibfnamefont {J.~B.}\ \bibnamefont
  {Pendry}},\ }\href {\doibase 10.1103/PhysRevLett.123.206101} {\bibfield
  {journal} {\bibinfo  {journal} {Physical Review Letters}\ }\textbf {\bibinfo
  {volume} {123}},\ \bibinfo {pages} {206101} (\bibinfo {year}
  {2019})}\BibitemShut {NoStop}%
\bibitem [{\citenamefont {Wang}\ \emph
  {et~al.}(2018{\natexlab{a}})\citenamefont {Wang}, \citenamefont
  {Yousefzadeh}, \citenamefont {Chen}, \citenamefont {Nassar}, \citenamefont
  {Huang},\ and\ \citenamefont {Daraio}}]{Wang2018}%
  \BibitemOpen
  \bibfield  {author} {\bibinfo {author} {\bibfnamefont {Y.}~\bibnamefont
  {Wang}}, \bibinfo {author} {\bibfnamefont {B.}~\bibnamefont {Yousefzadeh}},
  \bibinfo {author} {\bibfnamefont {H.}~\bibnamefont {Chen}}, \bibinfo {author}
  {\bibfnamefont {H.}~\bibnamefont {Nassar}}, \bibinfo {author} {\bibfnamefont
  {G.}~\bibnamefont {Huang}}, \ and\ \bibinfo {author} {\bibfnamefont
  {C.}~\bibnamefont {Daraio}},\ }\href {\doibase
  10.1103/PhysRevLett.121.194301} {\bibfield  {journal} {\bibinfo  {journal}
  {Physical Review Letters}\ }\textbf {\bibinfo {volume} {121}} (\bibinfo
  {year} {2018}{\natexlab{a}}),\ 10.1103/PhysRevLett.121.194301}\BibitemShut
  {NoStop}%
\bibitem [{\citenamefont {Trainiti}\ \emph {et~al.}(2019)\citenamefont
  {Trainiti}, \citenamefont {Xia}, \citenamefont {Marconi}, \citenamefont
  {Cazzulani}, \citenamefont {Erturk},\ and\ \citenamefont
  {Ruzzene}}]{Trainiti2019}%
  \BibitemOpen
  \bibfield  {author} {\bibinfo {author} {\bibfnamefont {G.}~\bibnamefont
  {Trainiti}}, \bibinfo {author} {\bibfnamefont {Y.}~\bibnamefont {Xia}},
  \bibinfo {author} {\bibfnamefont {J.}~\bibnamefont {Marconi}}, \bibinfo
  {author} {\bibfnamefont {G.}~\bibnamefont {Cazzulani}}, \bibinfo {author}
  {\bibfnamefont {A.}~\bibnamefont {Erturk}}, \ and\ \bibinfo {author}
  {\bibfnamefont {M.}~\bibnamefont {Ruzzene}},\ }\href {\doibase
  10.1103/PhysRevLett.122.124301} {\bibfield  {journal} {\bibinfo  {journal}
  {Physical Review Letters}\ }\textbf {\bibinfo {volume} {122}},\ \bibinfo
  {pages} {124301} (\bibinfo {year} {2019})}\BibitemShut {NoStop}%
\bibitem [{\citenamefont {Yi}\ \emph {et~al.}(2019)\citenamefont {Yi},
  \citenamefont {Ouisse}, \citenamefont {{Sadoulet-Reboul}},\ and\
  \citenamefont {Matten}}]{Yi2019}%
  \BibitemOpen
  \bibfield  {author} {\bibinfo {author} {\bibfnamefont {K.}~\bibnamefont
  {Yi}}, \bibinfo {author} {\bibfnamefont {M.}~\bibnamefont {Ouisse}}, \bibinfo
  {author} {\bibfnamefont {E.}~\bibnamefont {{Sadoulet-Reboul}}}, \ and\
  \bibinfo {author} {\bibfnamefont {G.}~\bibnamefont {Matten}},\ }\href
  {\doibase 10.1088/1361-665X/ab19dc} {\bibfield  {journal} {\bibinfo
  {journal} {Smart Materials and Structures}\ }\textbf {\bibinfo {volume}
  {28}},\ \bibinfo {pages} {065025} (\bibinfo {year} {2019})}\BibitemShut
  {NoStop}%
\bibitem [{\citenamefont {Chen}\ \emph
  {et~al.}(2019{\natexlab{a}})\citenamefont {Chen}, \citenamefont {Li},
  \citenamefont {Nassar}, \citenamefont {Norris}, \citenamefont {Daraio},\ and\
  \citenamefont {Huang}}]{Chen2019b}%
  \BibitemOpen
  \bibfield  {author} {\bibinfo {author} {\bibfnamefont {Y.}~\bibnamefont
  {Chen}}, \bibinfo {author} {\bibfnamefont {X.}~\bibnamefont {Li}}, \bibinfo
  {author} {\bibfnamefont {H.}~\bibnamefont {Nassar}}, \bibinfo {author}
  {\bibfnamefont {A.~N.}\ \bibnamefont {Norris}}, \bibinfo {author}
  {\bibfnamefont {C.}~\bibnamefont {Daraio}}, \ and\ \bibinfo {author}
  {\bibfnamefont {G.}~\bibnamefont {Huang}},\ }\href {\doibase
  10.1103/PhysRevApplied.11.064052} {\bibfield  {journal} {\bibinfo  {journal}
  {Physical Review Applied}\ }\textbf {\bibinfo {volume} {11}},\ \bibinfo
  {pages} {064052} (\bibinfo {year} {2019}{\natexlab{a}})}\BibitemShut
  {NoStop}%
\bibitem [{\citenamefont {Attarzadeh}\ \emph {et~al.}(2020)\citenamefont
  {Attarzadeh}, \citenamefont {Callanan},\ and\ \citenamefont
  {Nouh}}]{Attarzadeh2020}%
  \BibitemOpen
  \bibfield  {author} {\bibinfo {author} {\bibfnamefont {M.}~\bibnamefont
  {Attarzadeh}}, \bibinfo {author} {\bibfnamefont {J.}~\bibnamefont
  {Callanan}}, \ and\ \bibinfo {author} {\bibfnamefont {M.}~\bibnamefont
  {Nouh}},\ }\href {\doibase 10.1103/PhysRevApplied.13.021001} {\bibfield
  {journal} {\bibinfo  {journal} {Physical Review Applied}\ }\textbf {\bibinfo
  {volume} {13}},\ \bibinfo {pages} {021001} (\bibinfo {year}
  {2020})}\BibitemShut {NoStop}%
\bibitem [{\citenamefont {Marconi}\ \emph {et~al.}(2020)\citenamefont
  {Marconi}, \citenamefont {Riva}, \citenamefont {Di~Ronco}, \citenamefont
  {Cazzulani}, \citenamefont {Braghin},\ and\ \citenamefont
  {Ruzzene}}]{Marconi2020}%
  \BibitemOpen
  \bibfield  {author} {\bibinfo {author} {\bibfnamefont {J.}~\bibnamefont
  {Marconi}}, \bibinfo {author} {\bibfnamefont {E.}~\bibnamefont {Riva}},
  \bibinfo {author} {\bibfnamefont {M.}~\bibnamefont {Di~Ronco}}, \bibinfo
  {author} {\bibfnamefont {G.}~\bibnamefont {Cazzulani}}, \bibinfo {author}
  {\bibfnamefont {F.}~\bibnamefont {Braghin}}, \ and\ \bibinfo {author}
  {\bibfnamefont {M.}~\bibnamefont {Ruzzene}},\ }\href {\doibase
  10.1103/PhysRevApplied.13.031001} {\bibfield  {journal} {\bibinfo  {journal}
  {Physical Review Applied}\ }\textbf {\bibinfo {volume} {13}},\ \bibinfo
  {pages} {031001} (\bibinfo {year} {2020})}\BibitemShut {NoStop}%
\bibitem [{\citenamefont {Xia}\ \emph {et~al.}(2021)\citenamefont {Xia},
  \citenamefont {Riva}, \citenamefont {Rosa}, \citenamefont {Cazzulani},
  \citenamefont {Erturk}, \citenamefont {Braghin},\ and\ \citenamefont
  {Ruzzene}}]{Xia2021}%
  \BibitemOpen
  \bibfield  {author} {\bibinfo {author} {\bibfnamefont {Y.}~\bibnamefont
  {Xia}}, \bibinfo {author} {\bibfnamefont {E.}~\bibnamefont {Riva}}, \bibinfo
  {author} {\bibfnamefont {M.~I.~N.}\ \bibnamefont {Rosa}}, \bibinfo {author}
  {\bibfnamefont {G.}~\bibnamefont {Cazzulani}}, \bibinfo {author}
  {\bibfnamefont {A.}~\bibnamefont {Erturk}}, \bibinfo {author} {\bibfnamefont
  {F.}~\bibnamefont {Braghin}}, \ and\ \bibinfo {author} {\bibfnamefont
  {M.}~\bibnamefont {Ruzzene}},\ }\href {\doibase
  10.1103/PhysRevLett.126.095501} {\bibfield  {journal} {\bibinfo  {journal}
  {Physical Review Letters}\ }\textbf {\bibinfo {volume} {126}},\ \bibinfo
  {pages} {095501} (\bibinfo {year} {2021})}\BibitemShut {NoStop}%
\bibitem [{\citenamefont {Nassar}\ \emph {et~al.}(2020)\citenamefont {Nassar},
  \citenamefont {Yousefzadeh}, \citenamefont {Fleury}, \citenamefont {Ruzzene},
  \citenamefont {Al{\`u}}, \citenamefont {Daraio}, \citenamefont {Norris},
  \citenamefont {Huang},\ and\ \citenamefont {Haberman}}]{Nassar2020}%
  \BibitemOpen
  \bibfield  {author} {\bibinfo {author} {\bibfnamefont {H.}~\bibnamefont
  {Nassar}}, \bibinfo {author} {\bibfnamefont {B.}~\bibnamefont {Yousefzadeh}},
  \bibinfo {author} {\bibfnamefont {R.}~\bibnamefont {Fleury}}, \bibinfo
  {author} {\bibfnamefont {M.}~\bibnamefont {Ruzzene}}, \bibinfo {author}
  {\bibfnamefont {A.}~\bibnamefont {Al{\`u}}}, \bibinfo {author} {\bibfnamefont
  {C.}~\bibnamefont {Daraio}}, \bibinfo {author} {\bibfnamefont {A.~N.}\
  \bibnamefont {Norris}}, \bibinfo {author} {\bibfnamefont {G.}~\bibnamefont
  {Huang}}, \ and\ \bibinfo {author} {\bibfnamefont {M.~R.}\ \bibnamefont
  {Haberman}},\ }\href {\doibase 10.1038/s41578-020-0206-0} {\bibfield
  {journal} {\bibinfo  {journal} {Nature Reviews Materials}\ ,\ \bibinfo
  {pages} {1}} (\bibinfo {year} {2020})}\BibitemShut {NoStop}%
\bibitem [{\citenamefont {Roe}\ and\ \citenamefont {Boyd}(1959)}]{Roe1959}%
  \BibitemOpen
  \bibfield  {author} {\bibinfo {author} {\bibfnamefont {G.~M.}\ \bibnamefont
  {Roe}}\ and\ \bibinfo {author} {\bibfnamefont {M.~R.}\ \bibnamefont {Boyd}},\
  }\href {\doibase 10.1109/JRPROC.1959.287353} {\bibfield  {journal} {\bibinfo
  {journal} {Proceedings of the IRE}\ }\textbf {\bibinfo {volume} {47}},\
  \bibinfo {pages} {1213} (\bibinfo {year} {1959})}\BibitemShut {NoStop}%
\bibitem [{\citenamefont {Simon}(1960)}]{Simon1960}%
  \BibitemOpen
  \bibfield  {author} {\bibinfo {author} {\bibfnamefont {J.-C.}\ \bibnamefont
  {Simon}},\ }\href {\doibase 10.1109/TMTT.1960.1124657} {\bibfield  {journal}
  {\bibinfo  {journal} {IRE Transactions on Microwave Theory and Techniques}\
  }\textbf {\bibinfo {volume} {8}},\ \bibinfo {pages} {18} (\bibinfo {year}
  {1960})}\BibitemShut {NoStop}%
\bibitem [{\citenamefont {Hessel}\ and\ \citenamefont
  {Oliner}(1961)}]{Hessel1961}%
  \BibitemOpen
  \bibfield  {author} {\bibinfo {author} {\bibfnamefont {A.}~\bibnamefont
  {Hessel}}\ and\ \bibinfo {author} {\bibfnamefont {A.}~\bibnamefont
  {Oliner}},\ }\href {\doibase 10.1109/TMTT.1961.1125340} {\bibfield  {journal}
  {\bibinfo  {journal} {IRE Transactions on Microwave Theory and Techniques}\
  }\textbf {\bibinfo {volume} {9}},\ \bibinfo {pages} {337} (\bibinfo {year}
  {1961})}\BibitemShut {NoStop}%
\bibitem [{\citenamefont {Swinteck}\ \emph {et~al.}(2015)\citenamefont
  {Swinteck}, \citenamefont {Matsuo}, \citenamefont {Runge}, \citenamefont
  {Vasseur}, \citenamefont {Lucas},\ and\ \citenamefont
  {Deymier}}]{Swinteck2015}%
  \BibitemOpen
  \bibfield  {author} {\bibinfo {author} {\bibfnamefont {N.}~\bibnamefont
  {Swinteck}}, \bibinfo {author} {\bibfnamefont {S.}~\bibnamefont {Matsuo}},
  \bibinfo {author} {\bibfnamefont {K.}~\bibnamefont {Runge}}, \bibinfo
  {author} {\bibfnamefont {J.~O.}\ \bibnamefont {Vasseur}}, \bibinfo {author}
  {\bibfnamefont {P.}~\bibnamefont {Lucas}}, \ and\ \bibinfo {author}
  {\bibfnamefont {P.~A.}\ \bibnamefont {Deymier}},\ }\href {\doibase
  10.1063/1.4928619} {\bibfield  {journal} {\bibinfo  {journal} {Journal of
  Applied Physics}\ }\textbf {\bibinfo {volume} {118}},\ \bibinfo {pages}
  {063103} (\bibinfo {year} {2015})}\BibitemShut {NoStop}%
\bibitem [{\citenamefont {Trainiti}\ and\ \citenamefont
  {Ruzzene}(2016)}]{Trainiti2016}%
  \BibitemOpen
  \bibfield  {author} {\bibinfo {author} {\bibfnamefont {G.}~\bibnamefont
  {Trainiti}}\ and\ \bibinfo {author} {\bibfnamefont {M.}~\bibnamefont
  {Ruzzene}},\ }\href {\doibase 10.1088/1367-2630/18/8/083047} {\bibfield
  {journal} {\bibinfo  {journal} {New Journal of Physics}\ }\textbf {\bibinfo
  {volume} {18}},\ \bibinfo {pages} {083047} (\bibinfo {year}
  {2016})}\BibitemShut {NoStop}%
\bibitem [{\citenamefont {{Nassar H.}}\ \emph {et~al.}(2017)\citenamefont
  {{Nassar H.}}, \citenamefont {{Chen H.}}, \citenamefont {{Norris A. N.}},
  \citenamefont {{Haberman M. R.}},\ and\ \citenamefont {{Huang G.
  L.}}}]{NassarH.2017}%
  \BibitemOpen
  \bibfield  {author} {\bibinfo {author} {\bibnamefont {{Nassar H.}}}, \bibinfo
  {author} {\bibnamefont {{Chen H.}}}, \bibinfo {author} {\bibnamefont {{Norris
  A. N.}}}, \bibinfo {author} {\bibnamefont {{Haberman M. R.}}}, \ and\
  \bibinfo {author} {\bibnamefont {{Huang G. L.}}},\ }\href {\doibase
  10.1098/rspa.2017.0188} {\bibfield  {journal} {\bibinfo  {journal}
  {Proceedings of the Royal Society A: Mathematical, Physical and Engineering
  Sciences}\ }\textbf {\bibinfo {volume} {473}},\ \bibinfo {pages} {20170188}
  (\bibinfo {year} {2017})}\BibitemShut {NoStop}%
\bibitem [{\citenamefont {Wang}\ \emph
  {et~al.}(2018{\natexlab{b}})\citenamefont {Wang}, \citenamefont {Zhang},\
  and\ \citenamefont {Chan}}]{Wang2018a}%
  \BibitemOpen
  \bibfield  {author} {\bibinfo {author} {\bibfnamefont {N.}~\bibnamefont
  {Wang}}, \bibinfo {author} {\bibfnamefont {Z.-Q.}\ \bibnamefont {Zhang}}, \
  and\ \bibinfo {author} {\bibfnamefont {C.~T.}\ \bibnamefont {Chan}},\ }\href
  {\doibase 10.1103/PhysRevB.98.085142} {\bibfield  {journal} {\bibinfo
  {journal} {Physical Review B}\ }\textbf {\bibinfo {volume} {98}},\ \bibinfo
  {pages} {085142} (\bibinfo {year} {2018}{\natexlab{b}})}\BibitemShut
  {NoStop}%
\bibitem [{\citenamefont {Ashcroft}\ \emph {et~al.}(2016)\citenamefont
  {Ashcroft}, \citenamefont {Ashcroft}, \citenamefont {Wei}, \citenamefont
  {Mermin},\ and\ \citenamefont {Learning}}]{ashcroft2016solid}%
  \BibitemOpen
  \bibfield  {author} {\bibinfo {author} {\bibfnamefont {N.}~\bibnamefont
  {Ashcroft}}, \bibinfo {author} {\bibfnamefont {M.}~\bibnamefont {Ashcroft}},
  \bibinfo {author} {\bibfnamefont {D.}~\bibnamefont {Wei}}, \bibinfo {author}
  {\bibfnamefont {N.}~\bibnamefont {Mermin}}, \ and\ \bibinfo {author}
  {\bibfnamefont {C.}~\bibnamefont {Learning}},\ }\href@noop {} {\emph
  {\bibinfo {title} {Solid State Physics: Revised Edition}}}\ (\bibinfo
  {publisher} {{CENGAGE Learning Asia}},\ \bibinfo {year} {2016})\BibitemShut
  {NoStop}%
\bibitem [{\citenamefont {Matlack}\ \emph {et~al.}(2018)\citenamefont
  {Matlack}, \citenamefont {{Serra-Garcia}}, \citenamefont {Palermo},
  \citenamefont {Huber},\ and\ \citenamefont {Daraio}}]{Matlack2018a}%
  \BibitemOpen
  \bibfield  {author} {\bibinfo {author} {\bibfnamefont {K.~H.}\ \bibnamefont
  {Matlack}}, \bibinfo {author} {\bibfnamefont {M.}~\bibnamefont
  {{Serra-Garcia}}}, \bibinfo {author} {\bibfnamefont {A.}~\bibnamefont
  {Palermo}}, \bibinfo {author} {\bibfnamefont {S.~D.}\ \bibnamefont {Huber}},
  \ and\ \bibinfo {author} {\bibfnamefont {C.}~\bibnamefont {Daraio}},\ }\href
  {\doibase 10.1038/s41563-017-0003-3} {\bibfield  {journal} {\bibinfo
  {journal} {Nature Materials}\ }\textbf {\bibinfo {volume} {17}},\ \bibinfo
  {pages} {323} (\bibinfo {year} {2018})}\BibitemShut {NoStop}%
\bibitem [{\citenamefont {Karki}\ and\ \citenamefont
  {Paulose}(2021)}]{Karki2021}%
  \BibitemOpen
  \bibfield  {author} {\bibinfo {author} {\bibfnamefont {P.}~\bibnamefont
  {Karki}}\ and\ \bibinfo {author} {\bibfnamefont {J.}~\bibnamefont
  {Paulose}},\ }\href {\doibase 10.1103/PhysRevApplied.15.034083} {\bibfield
  {journal} {\bibinfo  {journal} {Physical Review Applied}\ }\textbf {\bibinfo
  {volume} {15}},\ \bibinfo {pages} {034083} (\bibinfo {year}
  {2021})}\BibitemShut {NoStop}%
\bibitem [{\citenamefont {Kovacic}\ \emph {et~al.}(2018)\citenamefont
  {Kovacic}, \citenamefont {Rand},\ and\ \citenamefont
  {Mohamed~Sah}}]{Kovacic2018}%
  \BibitemOpen
  \bibfield  {author} {\bibinfo {author} {\bibfnamefont {I.}~\bibnamefont
  {Kovacic}}, \bibinfo {author} {\bibfnamefont {R.}~\bibnamefont {Rand}}, \
  and\ \bibinfo {author} {\bibfnamefont {S.}~\bibnamefont {Mohamed~Sah}},\
  }\href {\doibase 10.1115/1.4039144} {\bibfield  {journal} {\bibinfo
  {journal} {Applied Mechanics Reviews}\ }\textbf {\bibinfo {volume} {70}}
  (\bibinfo {year} {2018}),\ 10.1115/1.4039144}\BibitemShut {NoStop}%
\bibitem [{\citenamefont {Chen}\ \emph
  {et~al.}(2019{\natexlab{b}})\citenamefont {Chen}, \citenamefont {Yao},
  \citenamefont {Nassar},\ and\ \citenamefont {Huang}}]{Chen2019a}%
  \BibitemOpen
  \bibfield  {author} {\bibinfo {author} {\bibfnamefont {H.}~\bibnamefont
  {Chen}}, \bibinfo {author} {\bibfnamefont {L.}~\bibnamefont {Yao}}, \bibinfo
  {author} {\bibfnamefont {H.}~\bibnamefont {Nassar}}, \ and\ \bibinfo {author}
  {\bibfnamefont {G.}~\bibnamefont {Huang}},\ }\href {\doibase
  10.1103/PhysRevApplied.11.044029} {\bibfield  {journal} {\bibinfo  {journal}
  {Physical Review Applied}\ }\textbf {\bibinfo {volume} {11}},\ \bibinfo
  {pages} {044029} (\bibinfo {year} {2019}{\natexlab{b}})}\BibitemShut
  {NoStop}%
\bibitem [{\citenamefont {Cha}\ and\ \citenamefont {Daraio}(2018)}]{Cha2018}%
  \BibitemOpen
  \bibfield  {author} {\bibinfo {author} {\bibfnamefont {J.}~\bibnamefont
  {Cha}}\ and\ \bibinfo {author} {\bibfnamefont {C.}~\bibnamefont {Daraio}},\
  }\href {\doibase 10.1038/s41565-018-0252-6} {\bibfield  {journal} {\bibinfo
  {journal} {Nature Nanotechnology}\ }\textbf {\bibinfo {volume} {13}},\
  \bibinfo {pages} {1016} (\bibinfo {year} {2018})}\BibitemShut {NoStop}%
\bibitem [{\citenamefont {Ghatak}\ \emph {et~al.}(2020)\citenamefont {Ghatak},
  \citenamefont {Brandenbourger}, \citenamefont {van Wezel},\ and\
  \citenamefont {Coulais}}]{Ghatak2020}%
  \BibitemOpen
  \bibfield  {author} {\bibinfo {author} {\bibfnamefont {A.}~\bibnamefont
  {Ghatak}}, \bibinfo {author} {\bibfnamefont {M.}~\bibnamefont
  {Brandenbourger}}, \bibinfo {author} {\bibfnamefont {J.}~\bibnamefont {van
  Wezel}}, \ and\ \bibinfo {author} {\bibfnamefont {C.}~\bibnamefont
  {Coulais}},\ }\href {\doibase 10.1073/pnas.2010580117} {\bibfield  {journal}
  {\bibinfo  {journal} {Proceedings of the National Academy of Sciences}\
  }\textbf {\bibinfo {volume} {117}},\ \bibinfo {pages} {29561} (\bibinfo
  {year} {2020})}\BibitemShut {NoStop}%
\bibitem [{\citenamefont {Rosa}\ and\ \citenamefont
  {Ruzzene}(2020)}]{Rosa2020}%
  \BibitemOpen
  \bibfield  {author} {\bibinfo {author} {\bibfnamefont {M.~I.~N.}\
  \bibnamefont {Rosa}}\ and\ \bibinfo {author} {\bibfnamefont {M.}~\bibnamefont
  {Ruzzene}},\ }\href {\doibase 10.1088/1367-2630/ab81b6} {\bibfield  {journal}
  {\bibinfo  {journal} {New Journal of Physics}\ }\textbf {\bibinfo {volume}
  {22}},\ \bibinfo {pages} {053004} (\bibinfo {year} {2020})}\BibitemShut
  {NoStop}%
\bibitem [{\citenamefont {Braghini}\ \emph {et~al.}(2021)\citenamefont
  {Braghini}, \citenamefont {Villani}, \citenamefont {Rosa},\ and\
  \citenamefont {Arruda}}]{Braghini2021}%
  \BibitemOpen
  \bibfield  {author} {\bibinfo {author} {\bibfnamefont {D.}~\bibnamefont
  {Braghini}}, \bibinfo {author} {\bibfnamefont {L.~G.~G.}\ \bibnamefont
  {Villani}}, \bibinfo {author} {\bibfnamefont {M.~I.~N.}\ \bibnamefont
  {Rosa}}, \ and\ \bibinfo {author} {\bibfnamefont {J.~R. d.~F.}\ \bibnamefont
  {Arruda}},\ }\href {\doibase 10.1088/1361-6463/abf9d9} {\bibfield  {journal}
  {\bibinfo  {journal} {Journal of Physics D: Applied Physics}\ }\textbf
  {\bibinfo {volume} {54}},\ \bibinfo {pages} {285302} (\bibinfo {year}
  {2021})}\BibitemShut {NoStop}%
\bibitem [{\citenamefont {Coulais}\ \emph {et~al.}(2021)\citenamefont
  {Coulais}, \citenamefont {Fleury},\ and\ \citenamefont {{van
  Wezel}}}]{Coulais2021}%
  \BibitemOpen
  \bibfield  {author} {\bibinfo {author} {\bibfnamefont {C.}~\bibnamefont
  {Coulais}}, \bibinfo {author} {\bibfnamefont {R.}~\bibnamefont {Fleury}}, \
  and\ \bibinfo {author} {\bibfnamefont {J.}~\bibnamefont {{van Wezel}}},\
  }\href {\doibase 10.1038/s41567-020-01093-z} {\bibfield  {journal} {\bibinfo
  {journal} {Nature Physics}\ }\textbf {\bibinfo {volume} {17}},\ \bibinfo
  {pages} {9} (\bibinfo {year} {2021})}\BibitemShut {NoStop}%
\bibitem [{\citenamefont {Salerno}\ \emph {et~al.}(2016)\citenamefont
  {Salerno}, \citenamefont {Ozawa}, \citenamefont {Price},\ and\ \citenamefont
  {Carusotto}}]{Salerno2016}%
  \BibitemOpen
  \bibfield  {author} {\bibinfo {author} {\bibfnamefont {G.}~\bibnamefont
  {Salerno}}, \bibinfo {author} {\bibfnamefont {T.}~\bibnamefont {Ozawa}},
  \bibinfo {author} {\bibfnamefont {H.~M.}\ \bibnamefont {Price}}, \ and\
  \bibinfo {author} {\bibfnamefont {I.}~\bibnamefont {Carusotto}},\ }\href
  {\doibase 10.1103/PhysRevB.93.085105} {\bibfield  {journal} {\bibinfo
  {journal} {Physical Review B}\ }\textbf {\bibinfo {volume} {93}},\ \bibinfo
  {pages} {085105} (\bibinfo {year} {2016})},\ \Eprint
  {http://arxiv.org/abs/1510.04697} {arXiv:1510.04697} \BibitemShut {NoStop}%
\bibitem [{\citenamefont {Zanjani}\ \emph {et~al.}(2014)\citenamefont
  {Zanjani}, \citenamefont {Davoyan}, \citenamefont {Mahmoud}, \citenamefont
  {Engheta},\ and\ \citenamefont {Lukes}}]{Zanjani2014}%
  \BibitemOpen
  \bibfield  {author} {\bibinfo {author} {\bibfnamefont {M.~B.}\ \bibnamefont
  {Zanjani}}, \bibinfo {author} {\bibfnamefont {A.~R.}\ \bibnamefont
  {Davoyan}}, \bibinfo {author} {\bibfnamefont {A.~M.}\ \bibnamefont
  {Mahmoud}}, \bibinfo {author} {\bibfnamefont {N.}~\bibnamefont {Engheta}}, \
  and\ \bibinfo {author} {\bibfnamefont {J.~R.}\ \bibnamefont {Lukes}},\ }\href
  {\doibase 10.1063/1.4866590} {\bibfield  {journal} {\bibinfo  {journal}
  {Applied Physics Letters}\ }\textbf {\bibinfo {volume} {104}},\ \bibinfo
  {pages} {081905} (\bibinfo {year} {2014})}\BibitemShut {NoStop}%
\bibitem [{\citenamefont {Nassar}\ \emph {et~al.}(2017)\citenamefont {Nassar},
  \citenamefont {Chen}, \citenamefont {Norris},\ and\ \citenamefont
  {Huang}}]{Nassar2017a}%
  \BibitemOpen
  \bibfield  {author} {\bibinfo {author} {\bibfnamefont {H.}~\bibnamefont
  {Nassar}}, \bibinfo {author} {\bibfnamefont {H.}~\bibnamefont {Chen}},
  \bibinfo {author} {\bibfnamefont {A.~N.}\ \bibnamefont {Norris}}, \ and\
  \bibinfo {author} {\bibfnamefont {G.~L.}\ \bibnamefont {Huang}},\ }\href
  {\doibase 10.1016/j.eml.2017.07.001} {\bibfield  {journal} {\bibinfo
  {journal} {Extreme Mechanics Letters}\ }\textbf {\bibinfo {volume} {15}},\
  \bibinfo {pages} {97} (\bibinfo {year} {2017})}\BibitemShut {NoStop}%
\bibitem [{\citenamefont {Deymier}\ \emph {et~al.}(2017)\citenamefont
  {Deymier}, \citenamefont {Gole}, \citenamefont {Lucas}, \citenamefont
  {Vasseur},\ and\ \citenamefont {Runge}}]{Deymier2017}%
  \BibitemOpen
  \bibfield  {author} {\bibinfo {author} {\bibfnamefont {P.~A.}\ \bibnamefont
  {Deymier}}, \bibinfo {author} {\bibfnamefont {V.}~\bibnamefont {Gole}},
  \bibinfo {author} {\bibfnamefont {P.}~\bibnamefont {Lucas}}, \bibinfo
  {author} {\bibfnamefont {J.~O.}\ \bibnamefont {Vasseur}}, \ and\ \bibinfo
  {author} {\bibfnamefont {K.}~\bibnamefont {Runge}},\ }\href {\doibase
  10.1103/PhysRevB.96.064304} {\bibfield  {journal} {\bibinfo  {journal}
  {Physical Review B}\ }\textbf {\bibinfo {volume} {96}},\ \bibinfo {pages}
  {064304} (\bibinfo {year} {2017})}\BibitemShut {NoStop}%
\bibitem [{\citenamefont {Vila}\ \emph {et~al.}(2017)\citenamefont {Vila},
  \citenamefont {Pal}, \citenamefont {Ruzzene},\ and\ \citenamefont
  {Trainiti}}]{Vila2017}%
  \BibitemOpen
  \bibfield  {author} {\bibinfo {author} {\bibfnamefont {J.}~\bibnamefont
  {Vila}}, \bibinfo {author} {\bibfnamefont {R.~K.}\ \bibnamefont {Pal}},
  \bibinfo {author} {\bibfnamefont {M.}~\bibnamefont {Ruzzene}}, \ and\
  \bibinfo {author} {\bibfnamefont {G.}~\bibnamefont {Trainiti}},\ }\href
  {\doibase 10.1016/j.jsv.2017.06.011} {\bibfield  {journal} {\bibinfo
  {journal} {Journal of Sound and Vibration}\ }\textbf {\bibinfo {volume}
  {406}},\ \bibinfo {pages} {363} (\bibinfo {year} {2017})}\BibitemShut
  {NoStop}%
\bibitem [{\citenamefont {Attarzadeh}\ and\ \citenamefont
  {Nouh}(2018)}]{Attarzadeh2018}%
  \BibitemOpen
  \bibfield  {author} {\bibinfo {author} {\bibfnamefont {M.~A.}\ \bibnamefont
  {Attarzadeh}}\ and\ \bibinfo {author} {\bibfnamefont {M.}~\bibnamefont
  {Nouh}},\ }\href {\doibase 10.1016/j.jsv.2018.02.028} {\bibfield  {journal}
  {\bibinfo  {journal} {Journal of Sound and Vibration}\ }\textbf {\bibinfo
  {volume} {422}},\ \bibinfo {pages} {264} (\bibinfo {year}
  {2018})}\BibitemShut {NoStop}%
\bibitem [{\citenamefont {Attarzadeh}\ \emph {et~al.}(2018)\citenamefont
  {Attarzadeh}, \citenamefont {Al~Ba'ba'a},\ and\ \citenamefont
  {Nouh}}]{Attarzadeh2018a}%
  \BibitemOpen
  \bibfield  {author} {\bibinfo {author} {\bibfnamefont {M.~A.}\ \bibnamefont
  {Attarzadeh}}, \bibinfo {author} {\bibfnamefont {H.}~\bibnamefont
  {Al~Ba'ba'a}}, \ and\ \bibinfo {author} {\bibfnamefont {M.}~\bibnamefont
  {Nouh}},\ }\href {\doibase 10.1016/j.apacoust.2017.12.028} {\bibfield
  {journal} {\bibinfo  {journal} {Applied Acoustics}\ }\textbf {\bibinfo
  {volume} {133}},\ \bibinfo {pages} {210} (\bibinfo {year}
  {2018})}\BibitemShut {NoStop}%
\bibitem [{\citenamefont {Nassar}\ \emph {et~al.}(2018)\citenamefont {Nassar},
  \citenamefont {Chen}, \citenamefont {Norris},\ and\ \citenamefont
  {Huang}}]{Nassar2018}%
  \BibitemOpen
  \bibfield  {author} {\bibinfo {author} {\bibfnamefont {H.}~\bibnamefont
  {Nassar}}, \bibinfo {author} {\bibfnamefont {H.}~\bibnamefont {Chen}},
  \bibinfo {author} {\bibfnamefont {A.~N.}\ \bibnamefont {Norris}}, \ and\
  \bibinfo {author} {\bibfnamefont {G.~L.}\ \bibnamefont {Huang}},\ }\href
  {\doibase 10.1103/PhysRevB.97.014305} {\bibfield  {journal} {\bibinfo
  {journal} {Physical Review B}\ }\textbf {\bibinfo {volume} {97}},\ \bibinfo
  {pages} {014305} (\bibinfo {year} {2018})}\BibitemShut {NoStop}%
\bibitem [{\citenamefont {Li}\ \emph {et~al.}(2019)\citenamefont {Li},
  \citenamefont {Ni}, \citenamefont {Weiner}, \citenamefont {Al{\`u}},\ and\
  \citenamefont {Khanikaev}}]{Li2019a}%
  \BibitemOpen
  \bibfield  {author} {\bibinfo {author} {\bibfnamefont {M.}~\bibnamefont
  {Li}}, \bibinfo {author} {\bibfnamefont {X.}~\bibnamefont {Ni}}, \bibinfo
  {author} {\bibfnamefont {M.}~\bibnamefont {Weiner}}, \bibinfo {author}
  {\bibfnamefont {A.}~\bibnamefont {Al{\`u}}}, \ and\ \bibinfo {author}
  {\bibfnamefont {A.~B.}\ \bibnamefont {Khanikaev}},\ }\href {\doibase
  10.1103/PhysRevB.100.045423} {\bibfield  {journal} {\bibinfo  {journal}
  {Physical Review B}\ }\textbf {\bibinfo {volume} {100}},\ \bibinfo {pages}
  {045423} (\bibinfo {year} {2019})}\BibitemShut {NoStop}%
\bibitem [{\citenamefont {{G{\'o}mez-Le{\'o}n}}\ and\ \citenamefont
  {Platero}(2013)}]{Gomez-Leon2013}%
  \BibitemOpen
  \bibfield  {author} {\bibinfo {author} {\bibfnamefont {a.}~\bibnamefont
  {{G{\'o}mez-Le{\'o}n}}}\ and\ \bibinfo {author} {\bibfnamefont
  {G.}~\bibnamefont {Platero}},\ }\href {\doibase
  10.1103/PhysRevLett.110.200403} {\bibfield  {journal} {\bibinfo  {journal}
  {Physical Review Letters}\ }\textbf {\bibinfo {volume} {110}},\ \bibinfo
  {pages} {1} (\bibinfo {year} {2013})},\ \Eprint
  {http://arxiv.org/abs/1303.4369} {arXiv:1303.4369} \BibitemShut {NoStop}%
\bibitem [{\citenamefont {Holthaus}(2015)}]{Holthaus2015}%
  \BibitemOpen
  \bibfield  {author} {\bibinfo {author} {\bibfnamefont {M.}~\bibnamefont
  {Holthaus}},\ }\href {\doibase 10.1088/0953-4075/49/1/013001} {\bibfield
  {journal} {\bibinfo  {journal} {Journal of Physics B: Atomic, Molecular and
  Optical Physics}\ }\textbf {\bibinfo {volume} {49}},\ \bibinfo {pages}
  {013001} (\bibinfo {year} {2015})}\BibitemShut {NoStop}%
\bibitem [{\citenamefont {Iakubovich}\ and\ \citenamefont
  {Starzhinskii}(1975)}]{Iakubovich1975}%
  \BibitemOpen
  \bibfield  {author} {\bibinfo {author} {\bibfnamefont {V.~A.}\ \bibnamefont
  {Iakubovich}}\ and\ \bibinfo {author} {\bibfnamefont {V.~M.}\ \bibnamefont
  {Starzhinskii}},\ }\href@noop {} {\emph {\bibinfo {title} {Linear
  differential equations with periodic coefficients}}}\ (\bibinfo  {publisher}
  {Wiley},\ \bibinfo {year} {1975})\BibitemShut {NoStop}%
\bibitem [{\citenamefont {Anderson}\ \emph {et~al.}(2008)\citenamefont
  {Anderson}, \citenamefont {Lorenz},\ and\ \citenamefont
  {Travesset}}]{anderson2008}%
  \BibitemOpen
  \bibfield  {author} {\bibinfo {author} {\bibfnamefont {J.~A.}\ \bibnamefont
  {Anderson}}, \bibinfo {author} {\bibfnamefont {C.~D.}\ \bibnamefont
  {Lorenz}}, \ and\ \bibinfo {author} {\bibfnamefont {A.}~\bibnamefont
  {Travesset}},\ }\href {\doibase 10.1016/j.jcp.2008.01.047} {\bibfield
  {journal} {\bibinfo  {journal} {Journal of Computational Physics}\ }\textbf
  {\bibinfo {volume} {227}},\ \bibinfo {pages} {5342} (\bibinfo {year}
  {2008})},\ \bibinfo {note} {hOOMD-blue feature: HOOMD-blue}\BibitemShut
  {NoStop}%
\bibitem [{\citenamefont {Glaser}\ \emph {et~al.}(2015)\citenamefont {Glaser},
  \citenamefont {Nguyen}, \citenamefont {Anderson}, \citenamefont {Liu},
  \citenamefont {Spiga}, \citenamefont {Millan}, \citenamefont {Morse},\ and\
  \citenamefont {Glotzer}}]{glaser2015}%
  \BibitemOpen
  \bibfield  {author} {\bibinfo {author} {\bibfnamefont {J.}~\bibnamefont
  {Glaser}}, \bibinfo {author} {\bibfnamefont {T.~D.}\ \bibnamefont {Nguyen}},
  \bibinfo {author} {\bibfnamefont {J.~A.}\ \bibnamefont {Anderson}}, \bibinfo
  {author} {\bibfnamefont {P.}~\bibnamefont {Liu}}, \bibinfo {author}
  {\bibfnamefont {F.}~\bibnamefont {Spiga}}, \bibinfo {author} {\bibfnamefont
  {J.~A.}\ \bibnamefont {Millan}}, \bibinfo {author} {\bibfnamefont {D.~C.}\
  \bibnamefont {Morse}}, \ and\ \bibinfo {author} {\bibfnamefont {S.~C.}\
  \bibnamefont {Glotzer}},\ }\href {\doibase 10.1016/j.cpc.2015.02.028}
  {\bibfield  {journal} {\bibinfo  {journal} {Computer Physics Communications}\
  }\textbf {\bibinfo {volume} {192}},\ \bibinfo {pages} {97} (\bibinfo {year}
  {2015})},\ \bibinfo {note} {hOOMD-blue feature: HOOMD-blue}\BibitemShut
  {NoStop}%
\end{thebibliography}
\end{document}